\documentclass[useAMS,usenatbib]{mn2e}
\pdfoutput=1
\usepackage[pdftex]{graphicx}
\usepackage{natbib}
\usepackage{hyperref}
\hypersetup{pdfpagemode=UseNone} 

\newcommand{\msun}{M$_{\odot}$}

\newcommand{\tmax}{$T_{\rm max}\,$}
\newcommand{\tviracc}{$T_{\rm vir,acc}\,$}

\newcommand{\aapr}{A\&ARv}
\newcommand{\apj}{ApJ}
\newcommand{\apjl}{ApJL}
\newcommand{\apjs}{ApJS}
\newcommand{\aap}{A \& A}

\newcommand{\mnras}{MNRAS}

\newcommand{\nat}{Nature}

\title[The impact of feedback on cosmological gas accretion]{The impact of feedback on cosmological gas accretion}
\author[D. R. Nelson et al.]{Dylan Nelson$^{1}$\thanks{E-mail: dnelson@cfa.harvard.edu}, 
Shy Genel$^{1}$, Mark Vogelsberger$^{2}$, Volker Springel$^{3,4}$, \newauthor Debora Sijacki$^{5}$, Paul Torrey$^{2,6}$, Lars Hernquist$^{1}$\\\\
$^{1}$Harvard-Smithsonian Center for Astrophysics, 60 Garden Street, Cambridge, MA, 02138, USA\\
$^{2}$Kavli Institute for Astrophysics and Space Research, Department of Physics, MIT, Cambridge, MA 02139, USA\\
$^{3}$Heidelberg Institute for Theoretical Studies, Schloss-Wolfsbrunnenweg 35, 69118 Heidelberg, Germany\\
$^{4}$Zentrum f\"{u}r Astronomie der Universit\"{a}t Heidelberg, ARI, M\"{o}nchhofstr. 12-14, 69120 Heidelberg, Germany\\
$^{5}${Institute of Astronomy and Kavli Institute for Cosmology, University of Cambridge, Madingley Road, Cambridge CB3 0HA, UK}\\
$^{6}${TAPIR, Mailcode 350-17, California Institute of Technology, Pasadena, CA 91125, USA}\\
}

\begin{document}

\maketitle

\begin{abstract}
We investigate how the way galaxies acquire their gas across cosmic time in cosmological hydrodynamic simulations is 
modified by a comprehensive physical model for baryonic feedback processes. To do so, we compare two 
simulations -- with and without feedback -- both evolved with the moving mesh code {\small AREPO}. The feedback runs 
implement the full physics model of the Illustris simulation project, including star formation driven galactic winds and 
energetic feedback from supermassive blackholes. We explore: 
(a) the accretion rate of material contributing to the net growth of galaxies and originating directly from 
the intergalactic medium, finding that feedback strongly suppresses the raw, as well as the net, inflow of this 
``smooth mode'' gas at all redshifts, regardless of the temperature history of newly acquired gas.
(b) At the virial radius the temperature and radial flux of inflowing gas is largely unaffected at $z\!=\!2$. 
However, the spherical covering fraction of inflowing gas at 0.25\,$r_{\rm vir}$ decreases substantially, from more than 
80\% to less than 50\%, while the rates of both inflow and outflow increase, indicative of recycling across this boundary.
(c) The fractional contribution of smooth accretion to the total accretion rate is lower in the simulation with 
feedback, by roughly a factor of two across all redshifts. Moreover, the smooth component of gas with a cold temperature 
history, is entirely suppressed in the feedback run at $z\!<\!1$.
(d) The amount of time taken by gas to cross from the virial radius to the galaxy -- the ``halo transit 
time'' -- increases in the presence of feedback by a factor of $\simeq$\,2\,-\,3, and is notably independent of halo 
mass. We discuss the possible implications of this invariance for theoretical models of hot halo gas cooling.
\end{abstract}



\section{Introduction}


Over the past decade, numerical simulations modelling the evolution of gas in a $\Lambda$CDM cosmology have made it 
clear that the process by which galaxies acquire their baryons across cosmic time evades a satisfactory understanding.
In the ``classical'' theory of collapse and virialisation, gas from the intergalactic medium shock-heats to the virial 
temperature of a dark matter halo, subsequently forming a hot, pressure supported atmosphere in approximate equilibrium 
\citep{rees77,silk77,wr78}. The timescale of energy loss from radiative cooling determines the rate at which gas can 
cool into the halo centre \citep{wf91}, and indeed whether or not a stable virial shock can exist at all \citep{bd03}.

Numerical simulations over the past decade \citep[starting with][]{katz03,abadi03,keres05} have shown that gas 
accretion in the cosmological context, without the luxury of spherical symmetry, is significantly more complex. They 
found that (i) coherent streams of gas can provide strong fuelling for star formation over relatively small solid angles 
of the virial sphere, and (ii) such streams could potentially remain cold and avoid heating up to the virial temperature, 
even for haloes massive enough to support a quasi-static hot atmosphere. Such flows are a natural consequence of the 
``cosmic web'' of large scale structure, particularly at high redshift ($z\!>\!2$), and provide an intriguing avenue 
for gas accretion distinct from classic hot halo cooling. Aspherical gas inflow has been connected to many key questions 
in galaxy formation, including the growth of the stellar populations of galaxies \citep[e.g.][]{opp10}, their 
morphological transformations \citep{sales12,cen14}, kinematics \citep{genel12}, and star formation properties \citep{gabor14,almeida14}. 
It has also been looked at in terms of the acquisition of angular momentum \citep{danovich12,stewart13,danovich14}, 
including the connection to preferred directions imposed by the filaments of large scale structure \citep{dubois14}, 
and the feeding of supermassive blackhole accretion \citep{dubois12a,bellovary13,feng14}.


Smoothed particle hydrodynamics (SPH) simulations were the first to address the question of gas accretion modes in a 
cosmological context. Leveraging the quasi-Lagrangian nature of the numerical scheme, they could use the past temperature 
history of each gas element to differentiate between hot and cold mode accretion. It was found that the maximum past 
temperature of smoothly accreted gas is bimodal, the dominant contribution arising from gas which has never experienced 
significant heating during infall to the galaxy \citep{keres05}. 

However, caution is warranted. 
In our previous study \citep{nelson13} we compared the outcome of ``classic'' SPH simulations and those run with the 
newer moving mesh code {\small AREPO}. At $z\!=\!2$ we found a decrease in the accretion rate of cold gas, by a factor of 
$\sim 2$ at $M_{\rm halo} \simeq 10^{11}$\,\msun. We also found, at this same mass, an order of magnitude larger accretion 
rate of gas with significant past heating. These discrepancies grew even more significant for more massive haloes. 
We attributed the drop in the cold accretion rate to a large population of numerical ``blobs'' \citep{torrey12} which 
efficiently deliver cold gas to central galaxies in the SPH simulations, but are completely absent 
in the moving mesh calculations. The increase in the hot accretion rate was dominated by more efficient cooling from halo 
gas in {\small AREPO}, where spurious heating from the dissipation of turbulent energy on large scales prevents the 
correct behaviour in SPH \citep{bauer12}. Filamentary flows in the {\small AREPO} haloes were found to be warmer and more 
diffuse, and did not generally persist as strongly to small radii. 

Grid-based adaptive mesh refinement (AMR) simulations have also been interpreted as being in agreement with respect to the 
importance of a filamentary, cold accretion mode in massive systems at high redshift \citep{ocvirk08,dekel09,agertz09}, 
albeit with two potentially important limitations. In particular, these studies have often not used any form of Lagrangian 
tracer to follow the thermal and dynamical history of accreting gas, which would permit a more direct comparison with 
particle hydrodynamics codes. With the notable exception of \cite{ocvirk08}, they have also generally focused on targeted, 
''zoom'' simulations of individual haloes, where the significant halo-to-halo variation between codes \citep[e.g.][]{vog12} 
can make broadly applicable conclusions difficult. Regardless of hydrodynamical method, robust conclusions are difficult 
to draw from a single simulated halo, or a small sample of such haloes.

These findings made it clear that numerical deficiencies in the standard formulation of SPH used in past studies 
significantly biased previous quantitative conclusions as to the relative importance of cold streams or cold mode 
accretion \citep[also problematic are issues with formal numerical convergence, as discussed in][]{zhu14}. 
However, in this previous comparison work we included only a simple model for baryonic physics, appropriate to 
make an even-handed comparison to past work, but lacking the physical fidelity of modern cosmological simulations. The 
large question left outstanding was then: what impact, if any, does feedback associated with galaxy formation have on 
the process of gas accretion. Current state-of-the-art cosmological simulations have reached the point where they can 
evolve a significant volume of the universe down to $z\!=\!0$, while simultaneously resolving the structure 
of individual galaxies, and reproducing a broad range of observational constraints \citep{khandai14,vog14a,schaye14}. 
One of the many investigations related to galaxy formation and evolution that they enable is a study of baryonic accretion.


At the outset, it would seem entirely plausible that the accretion rates and the inflow of gas from the intergalactic 
medium would be quite sensitive to feedback processes. The impact could be either direct or indirect, or both. For 
instance, inflowing streams could be disrupted by spatially coincident outflows, such that the net mass flux entirely 
reversed direction. Or, energy injection from feedback could heat up the the surrounding hot halo gas, leading to a 
modification of the thermal history of inflow due to mixing. Alternatively, it would also seem plausible that feedback 
could have relatively little effect. For instance, star formation driven winds with non-isotropic outflow may simply 
evolve to occupy different regions of the virial volume than inflowing streams.

Recently, simulations have begun to investigate the additional complexity when feedback and galactic scale outflows are 
included \citep[beginning with][]{opp10,fgk11,vdv11a}. Conclusions as to their impact have been somewhat mixed, which 
undoubtedly arise from a combination of different feedback implementations, numerical methods, contexts, and interpretations. 
In brief review, 
\cite{brooks09} simulated five haloes, included a delayed cooling, supernova blastwave feedback model, but did not 
explicitly consider the impact of the feedback in {\small GASOLINE}, and in general found results consistent with \cite{keres05}.
\cite{opp10} included a kinetic galactic wind model in cosmological SPH simulations and concluded that 
recycled gas accretion is in fact the dominant accretion mechanism at $z\!\leq\!1$, with minimal effect on high 
redshift accretion.
\cite{fgk11} included a constant velocity galactic wind model, finding that net accretion rates measured as 
instantaneous mass fluxes could be substantially affected.
\cite{vdv11a} found that while the gas accretion rates onto haloes was relatively robust against the presence of 
feedback, the rates onto galaxies themselves depended sensitively on stellar winds as well as metal-line cooling.
\cite{vdv11b} studied the impact of AGN feedback on inflow and found that it preferentially prevented hot mode gas, 
with high maximum past temperature, from cooling from the halo onto the galaxy.
\cite{stewart11} simulates two relatively massive haloes, including the supernova blastwave feedback model, but notes 
this has little impact at the simulated mass scale, and does not consider how it modifies gas accretion nor AGN effects.
\cite{dubois12b} investigated AGN feedback at high redshift in one halo, and found that large-scale hot superwinds could 
morphologically disturb cold filaments and quench cold diffuse accretion.
\cite{murante12} used a thermal supernova feedback scheme, and found that additional heating of cold inflow due to this 
feedback gave rise to a significant accretion rate of intermediate temperature gas.
Most recently, \cite{woods14} includes a combined delayed cooling supernova and early stellar feedback model, finding that overall 
gas accretion rates did not change with strong feedback, while the balance between cold and hot components did.
Finally, \cite{ubler14} implements a hybrid thermal/kinetic stellar feedback scheme and finds strong outflows generate 
substantially higher raw accretion rates, and that recycled material dominates galactic gas accretion at $z\!<\!1$.

As a caveat, we note that the same numerical issues explored in the ``moving mesh cosmology'' series 
\citep{vog12,sijacki12,keres12,nelson13} that compromised the accuracy of SPH studies of gas inflow will also affect 
the interaction of outflowing ejecta and wind material with both halo and filamentary gas. 
Further, because hot gas in our {\small AREPO} simulations cools more efficiently than in classical SPH, the energy 
input from feedback required to prevent over-cooling is even larger than in previous simulations, implying that 
the impact on gas accretion could be significantly altered. We are therefore motivated to extend previous investigations 
with the current study, which combines a comprehensive, validated feedback model with an accurate and robust numerical 
technique in a systematic comparison.


This paper contrasts two simulations, realisations of the same initial conditions 
evolved with the moving mesh code {\small AREPO}. We compare populations of haloes and galaxies across cosmic time, 
contrasting the state and history of accreting gas between two runs, with and without feedback. In Section 
\ref{sMethods} we describe the simulation technique and analysis methodology. Section \ref{sOne} addresses the rate and 
history of primordial gas accretion, while Section \ref{sTwo} compares the instantaneous state of gas in haloes. 
Section \ref{sThree} considers the timescale of accretion through the halo. Finally, Sections \ref{sDiscussion} and 
\ref{sConclusions} discuss our results and their implications, and summarise our conclusions.


\section{Methods} \label{sMethods}

\subsection{The Simulations} \label{ssSimulations}

In this work we compare two simulations, ``with'' and ``without'' feedback, which have several common features. 
Both employ the {\small AREPO} code \citep{spr10} to solve the problem of ideal continuum hydrodynamics coupled with 
self-gravity. An unstructured, Voronoi tessellation of the simulation domain provides a spatial discretization for 
Godunov's method with a directionally un-split MUSCL-Hancock scheme \citep{vl77} and an exact Riemann solver, yielding 
second order accuracy in space. Since the mesh generating sites can be allowed to move, herein with a velocity 
tied to the local fluid velocity modulo mesh-regularization corrections, this numerical approach falls under the Arbitrary 
Lagrangian-Eulerian (ALE) class.
Gravitational forces are handled with the split Tree-PM approach, whereby long-range forces are calculated with a 
Fourier particle-mesh method, medium-range forces with a hierarchical tree algorithm \citep{bh86}, and short-range 
forces with direct summation.
A local, predictor-corrector type, hierarchical time stepping method yields second order accuracy in time.
Numerical parameters secondary to our current investigation -- for example, related to mesh regularization or 
gravitational force accuracy -- are detailed in \cite{spr10} and \cite{vog12}, and are unchanged between the two 
simulation sets.

Both simulations evolve the same initial condition, a random realisation of a WMAP-7 consistent cosmology 
($\Omega_{\Lambda,0}=0.73$, $\Omega_{m,0}=0.27$, $\Omega_{b,0}=0.045$, $\sigma_8=0.8$ and $h=0.7$) in a periodic 
cube of side-length 20$h^{-1}$ Mpc $\simeq 28.6$ Mpc, from a starting redshift of $z\!=\!99$ down to $z\!=\!0$. Each 
includes $512^3$ dark matter particles, an equal number of initial gas cells, and a minimum of $5 \times 512^3$ tracers 
(discussed below). The mean baryon mass is $1.1 \times 10^6$\,\msun, and the dark matter particle mass is 
$5.3 \times 10^6$\,\msun. The Plummer equivalent comoving gravitational softening lengths for dark matter and stars 
are $1.4$ kpc, and gas cells have adaptive softening lengths equal to 2.5 times their volume-equivalent spherical radius.
A redshift-dependent, spatially uniform, ionizing UV background field \citep{fg09} is included 
as a heating source. Star formation and the associated ISM pressurisation from unresolved supernovae are included 
with an effective equation of state modelling the ISM as a two-phase medium, following \cite{spr03}. Gas elements 
are stochastically converted into star particles when the local gas density exceeds a threshold value of 
$n_{\rm H}=0.13$ cm$^{-3}$. All of the simulations considered in this work disregard the possible effects of radiative 
transfer, magnetic fields, and cosmic rays.

The \textbf{no feedback runs} (``noFB'') with ``simple physics'' additionally account for optically thin radiative 
cooling assuming a primordial H/He ratio \citep{katz96}. They do not include metal line cooling, any resolved 
stellar feedback that would drive galactic-scale wind, nor any treatment of black holes or their associated feedback. 
This is the same simulation presented in \cite{nelson13}, where it was used in comparison to {\small GADGET} (SPH) results.

The \textbf{feedback runs} (``FB'') implement, unchanged, the fiducial physical model and associated parameter values of 
the Illustris simulation \citep{vog14a,vog14b,genel14} applied to the same initial conditions as our previous work, 
allowing object by object comparison. Complete details of the physics included in the model, as well 
as its tuning and validation, are described in \cite{vog13} and \cite{torrey14}. We describe here in some detail those 
aspects of the model which most strongly influence gas accretion.

First, we include the radiative cooling contribution from metal lines, where heavy elements are produced from supernovae 
Ia/II and AGB stars in stellar population evolution modelling \citep{thiel86,portinari98,karakas10}. In the absence of 
additional heating sources, this can increase the cooling rate $\Lambda$($n,T,Z,\Gamma$), for solar metallicity by an 
order of magnitude between $10^{4.5}$\,K\,$ < T_{\rm gas} < 10^{6.5}$\,K, mainly due to the contribution of O, Ne and Fe 
\citep{sd93,wiersma09}, enhancing cooling from the hot halo and the buildup of stellar mass.

To balance efficient cooling in the simulations, further exacerbated by metals, we include energetic feedback from star 
formation driven winds as well as supermassive black holes -- dominant in haloes less and more massive than M$_\star$, 
respectively. Stellar winds are generated directly from star-forming gas, with velocity $v_w = 3.7 \sigma_{\rm dm}$ where 
$\sigma_{\rm dm}$ is the local 1D dark matter velocity dispersion, which scales with the circular velocity maxima and so 
host (sub)halo total mass. In practice, gas cells are probabilistically converted into a wind-phase gas cell/particle, 
which interacts gravitationally but not hydrodynamically, until it reaches either a density threshold or a maximum 
travel time. Specifically, 0.05 times the star formation threshold in density, or 0.025 times the current Hubble time. 
This typically occurs just outside the disk -- that is, deep within the halo. At this point, the mass, momentum, metals, 
and internal energy are deposited into the gas cell in which the wind particle is located.

The energy-driven wind has a mass loading factor $\eta_w \propto v_w^{-2}$ (thereby decreasing 
with halo mass) and is assigned a ``bipolar'' outflow direction, given by the cross product of its original velocity and 
the gradient of the local potential. Since the wind speed is in general slightly less than the escape speed, this 
implementation can generate a strong ``galactic fountain'' effect of recycled material returning to the galaxy 
\citep{opp10,dave12}. As a result, it can also strongly modify the temperature and velocity structure of gas in the 
inner halo.

The second form of feedback originates in supermassive black holes \citep[SMBHs, following][]{sijacki07}, which are 
seeded with a mass of $1.4 \times 10^5$\,\msun\, in massive haloes above $7 \times 10^{10}$\,\msun\, and are effectively 
sink particles which grow through gas accretion and merging. When accretion onto 
the black hole is below $0.05$ of the Eddington rate, a radio-mode model injects highly bursty thermal energy equal to 
0.07 of the accreted rest mass energy in large ($\simeq$\,50 kpc) bubbles. When the accretion rate is above this fraction, 
a quasar-mode model injects thermal energy into nearby gas cells, with a lower coupling efficiency of 0.01 and 
a smoother time profile. For accretion rates approaching Eddington, a third, radiative form of feedback modifies the 
cooling rate for gas in the vicinity of the BH, assuming an optically-thin $1/r^2$ attenuation. In the halo mass 
regime where BH feedback becomes important, the radio-mode channel generates high velocity, high temperature outflows 
which influence gas at larger radii than the stellar feedback driven winds -- out to the virial radius and into the 
intergalactic medium.

There are three other minor changes with respect to the noFB configuration, required to match the fiducial Illustris 
model. First, we modify the equation of state parameter \citep{spr03} from q\,=\,1 to q\,=\,0.3, which interpolates between 
the effective EOS and an isothermal EOS of $10^4$\,K with weights of 0.3 and 0.7, respectively, in order to avoid over 
pressurising the ISM of star-forming gas. Secondly, we decouple 
the comoving gravitational softening lengths of the gas and stars from that of the dark matter at $z\!=\!1$, allowing the 
former to decrease by a further factor of two down to $z\!=\!0$. Finally, we include a correction for self-shielding of 
dense gas from the UV background \citep{rahmati13}.

\subsection{Monte Carlo Tracers} \label{ssMonteCarloTracers}

Both simulations include our new ``Monte Carlo tracer particle'' technique \citep{genel13} in order to follow 
the evolving properties of gas elements over time. This is a probabilistic method, where tracers act as unique tags 
in association with parent cells or particles. They have no phase space coordinates, but are instead exchanged between 
parents based explicitly on the corresponding mass fluxes. By locating a subset of their unique IDs at each snapshot 
we can, by reference to the gas cells in which the tracers reside (their parents), reconstruct their spatial trajectory or thermodynamic 
history. Furthermore, at each computational timestep every active tracer updates a record of its maximum temperature, 
density, Mach number and entropy, as well as the time of these events, enabling us to investigate these values with 
timestep-level resolution.

We extend the Monte Carlo tracer approach to include mass transfer between all baryonic components present 
in the simulations in a fully self-consistent manner. That is, tracers can reside in gas cells, star particles, 
wind-phase cells, and black holes, and exchange between these components in exactly the same ways that baryonic mass is 
exchanged during the simulation. In particular: 

\begin{enumerate}
\item Gas cell to gas cell transfer via finite volume fluxes, refinement and derefinement.
\item Gas cell to star particle via star formation, and the reverse during stellar mass return.
\item Gas cell to/from wind-phase during the generation/recoupling of star formation driven galactic winds.
\item Gas cell to black hole as a result of BH accretion, and between two black holes during a merger.
\end{enumerate}

This allows us to follow the flow of mass through all baryonic phases which are present. Finally, to explore the role 
of stellar feedback driven winds and recycling, each tracer also records the last time of exchange to a star/wind 
particle from gas, and the reverse, as well as a counter of the number of times it has been incorporated into a wind.

\subsection{Post-processing} \label{ssPostProcessing}

We identify dark matter haloes and their gravitationally bound substructures using the {\small SUBFIND} algorithm 
\citep{spr01,dolag09} which is applied on top of a friends-of-friends cluster identification. We refer to the most 
massive substructure in each FoF group as the halo itself, and consider accretion onto such haloes and the central 
galaxies hosted therein. We follow the evolving positions and properties of haloes over time by constructing a basic 
merger tree as in \cite{nelson13}, where only the ``main progenitor branch'' (MPB) is needed. For each halo, we 
restrict our analysis of accretion to the time period over which this main branch is robustly determined.

For each of a finite number of analysis redshifts, spanning $z\!=\!0$ to $z\!=\!5$, we perform a set of independent, 
identical analysis tasks. For all tracers (of all parent types) in all haloes at that redshift, we walk backwards and 
record the most recent time and direction of several particular radial crossings. 
We take $r_{\rm vir} = r_{\rm 200,crit}$ the radius enclosing a mean overdensity 200 times the critical density. 
We label the inward crossing times of two important radii, 0.15\,$r_{\rm vir}$ (representative of the outer boundary of the 
galaxy), and 1.0\,$r_{\rm vir}$ (representative of the outer boundary of the halo) the ``most recent incorporation time'' 
into the galaxy and the ``most recent accretion time'' into the halo, respectively. For these same two radii, we also 
record the earliest -- that is, highest redshift -- such crossing. We label the 0.15\,$r_{\rm vir}$ crossing as the 
``first incorporation time'', and the 1.0\,$r_{\rm vir}$ crossing as the ``first accretion time''. We discuss the 
calculation of accretion rates based on these quantities in the following section. 

Each tracer also records the virial temperature of its parent halo at the time of first 
accretion, labelled $T_{\rm vir,acc}$. We compare this value against $T_{\rm max}$, the maximum temperature of a tracer 
between the start of the simulation and the time of its first incorporation. This corresponds to the time at which 
each tracer accretes into the direct main progenitor of the central galaxy of the halo. As a result, the procedure is 
sensitive to virial shock heating in the MPB, as well as virial shock and feedback related heating in satellites prior 
to first incorporation, but not heating due to wind recycling after incorporation into the central galaxy.

In addition, we also separate all accreted tracers into one of three disjoint ``modes'' of 
accretion: smooth, clumpy, or stripped, according to the following definitions applied at the time of first accretion. 
Smooth: not a member of any resolved substructure, other than a MPB halo, and likewise at all previous times. Clumpy:  
gravitationally bound to any resolved substructure which is not a MPB halo. Stripped: otherwise smooth, but  
gravitationally bound to some resolved substructure other than the MBP at any previous time. In addition to these three 
modes, each tracer is also flagged as being ``recycled'' if it has ever been part of a stellar feedback driven wind at 
some previous time.

We note that identification of the ``clumpy'' component -- that is, the merger contribution -- includes only 
\textit{resolved} substructure. The combination of simulation resolution and group finder implies a minimum halo mass 
which can be identified. For the conclusions presented herein, this is M$_{\rm halo} \ge 10^{7.8}$\,\msun, meaning 
that for the halo mass range we consider all mergers with mass ratio above $\simeq 10^{-3}$ are correctly 
identified as a clumpy mode. The contribution of unresolved mergers below this level is expected to be negligible 
\citep[see][]{keres05,genel09}, and we have further verified that the associated conclusions are qualitatively unchanged 
in an identical simulation with a factor of eight lower mass resolution.


\section{Rate and Mode of Gas Accretion} \label{sOne}

In this section we compare the rate and mode of cosmological gas infall across cosmic time between the two 
simulations, with and without our fiducial feedback model. The presence of star formation driven galactic winds and 
AGN feedback, particularly in the radio-mode, generates significant radial velocity in the baryonic component of 
haloes. Outflows with high outward bulk motion in turn trigger ``galactic fountain'' behaviour, with significant gas 
mass returning after some time delay with high inward radial velocity. This efficient recycling through the galaxy 
itself implies that any instantaneous or quasi-instantaneous measurement of the total accretion rate of material 
into the galaxy will be a (possibly large) overestimate of the ``net'' rate of material permanently joining the 
galaxy by, for example, forming stars. Furthermore, this recycled material is a source of accretion onto galaxies 
in addition to direct cosmological accretion, which does not undergo a similar phase of processing and metal 
enrichment.

We are first interested in the question of whether the presence of strong galactic-scale feedback modifies the 
rate or character of cosmological -- or ``primordial'' -- infall. Therefore, to make a sensible comparison between 
simulations with and without the additional motions induced by outflows, we adopt the following approach. At any given 
redshift, accretion rates are measured as a tracer flux over a specific time interval, restricted to those 
baryons which are entering or leaving the galaxy, or its parent along the main progenitor branch, for the first time. 
We fix this time interval to be 250 Myr, but our results are not sensitive to this choice, so long as it is not too 
short to run up against the finite snapshot spacing, and not too long as to effectively smooth over changing physical 
conditions. In practice, we calculate three rates related to baryonic fluxes with respect to the galaxy:

\begin{enumerate}
\item
The outflow rate as the number of tracers with outward crossings through 0.15\,r$_{\rm vir}$ during this time window.
\item
The raw (as opposed to net) inflow rate as the number of tracers with inward crossings across this 
same boundary during this time window.
\item
The net accretion rate as the outflow rate subtracted from the inflow rate. That is, the difference of the number of 
tracers with inward 0.15\,r$_{\rm vir}$ crossing times and outward 0.15\,r$_{\rm vir}$ crossing times during this 
time window.
\end{enumerate}

\begin{figure}
\centerline{\includegraphics[angle=0,width=3.6in]{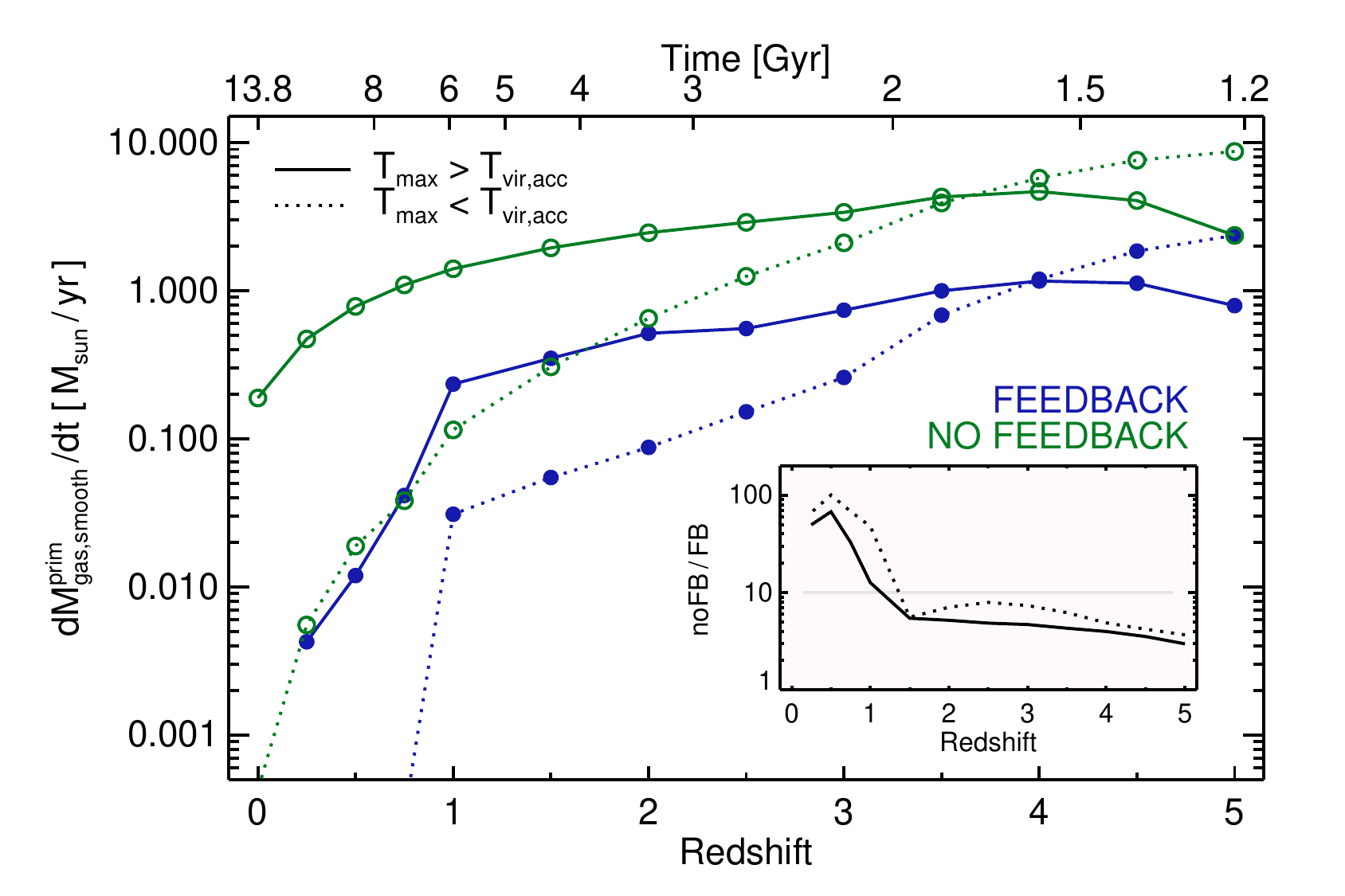}}
\caption{ The net smooth gas accretion rate of cosmological origin onto central galaxies as a function of redshift. 
At each redshift accretion over a time window of $\Delta t = 250 \,\rm{Myr}$ contributes. We include here only haloes 
in the mass range $11.3 < \log{(M_{\rm halo,tot}/\rm{M}_\odot)} < 11.4$, and separate the contribution based on the comparison of 
the maximum past temperature \tmax of each Monte Carlo tracer to the virial temperature of future parent halo, at the 
time of accretion. Both simulations, with (blue) and without (green) our fiducial feedback implementation, indicate 
that relatively cold gas dominates the primordial infall onto galaxies only at redshifts $\ga 4$, while at later times 
the reverse is true. At this mass scale, the impact of feedback is reduce the total rate of smooth cosmological 
accretion independent of the \tmax comparison, by a redshift-dependent amount (see inset).
 \label{fig_netrate_vs_redshift}} 
\end{figure}

In each case the tracer count is then multiplied by the tracer effective mass (the initial gas cell mass divided by 
the initial number of tracers per cell) and normalized by the time window to derive a rate. This method gives a measurement 
of $\dot{M}_{\rm gas}^{\rm prim} (M_{\rm halo}, z, \rm{dir}, \rm{mode})$ for $M_{\rm halo}$\,$\in$\,$[10^9,10^{12}]$\,\msun, 
$z \in [0,5]$, $\rm{dir} \in \{\rm{in,out,net}\}$ and $\rm{mode} \in \{\rm{all,smooth,clumpy,stripped,recycled}\}$ 
where ``prim'' denotes the primordial qualification. Alternatively, in Section \ref{ssArm0} we discuss 
another approach where the requirement on first incorporation is relaxed; instead we can calculate fluxes using 
the most recent incorporation time, thereby measuring a sum of material with two origins -- primordial as well as having 
previously cycled through the MPB. 

\subsection{Galactic accretion as a function of redshift}

In Fig. \ref{fig_netrate_vs_redshift} we compare the net smooth accretion rate between the two simulations, restricted to 
haloes in the mass range $11.3 < \log{(M_{\rm halo,tot}/\rm{M}_\odot)} < 11.4$ which we keep constant as a function of 
redshift. At $z\!=\!5$ the virial temperature of these haloes is $\simeq 10^6$\,K, decreasing to $\simeq 10^{5.3}$\,K 
at $z\!=\!0$. In this box, the mass selection contains 3 haloes at $z\!=\!5$, 42 at $z\!=\!2$, and 53 at $z\!=\!0$. 
Comparing the two simulations, we find that feedback suppresses smooth cosmological accretion by a factor 
of $\simeq$3 at high redshift ($z\!=\!5$). This increases to $\simeq$10 by $z\!=\!1$ after which the suppression continues 
to increase towards redshift zero, while the individual net rates also steadily decline. Indeed, the net smooth 
accretion rates drop by roughly an order of magnitude from $z\!=\!1$ to $z\!=\!0$ even in the noFB run, and this drop is 
also evident if we consider accretion over all modes, not just smooth. 
The star formation rate (SFR) of haloes in this mass range declines only moderately over this same time 
\citep[by a factor of $\sim$2,][]{genel14}, implying that the component of the late time SFR in these systems supplied 
by smooth gas accretion is supported predominantly by material which has cycled through the direct progenitor at an earlier time. 

We further split the smooth total based on the comparison between \tmax and \tviracc, available on a per tracer 
basis \citep[as in][]{nelson13}. In this mass regime, the balance of these two components is similar between both runs, 
where at $z\!>\!4$ gas with \tmax $<$ \tviracc dominates the smooth cosmological accretion budget, by up to a factor of 
three at $z\!=\!5$. Towards lower redshift gas with \tmax $>$ \tviracc instead dominates, by a factor of $\sim$4 at $z\!=\!2$ 
and by several orders of magnitude by $z\!=\!0$. As we discuss later in more detail, the similar balance of these two 
temperature components between the FB and noFB runs implies that the winds have marginal impact on the temperature 
history of smoothly accreting material, and that the presence of winds does not preferentially prevent material of a 
particular temperature history. To give a sense of reference, the difference between the mean \tmax of cold versus hot
temperature history gas is $\simeq$\,0.9 dex (at $z\!=\!5$) and $\simeq$\,0.6 dex (at $z\!=\!0$).

\begin{figure}
\centerline{\includegraphics[angle=0,width=3.6in]{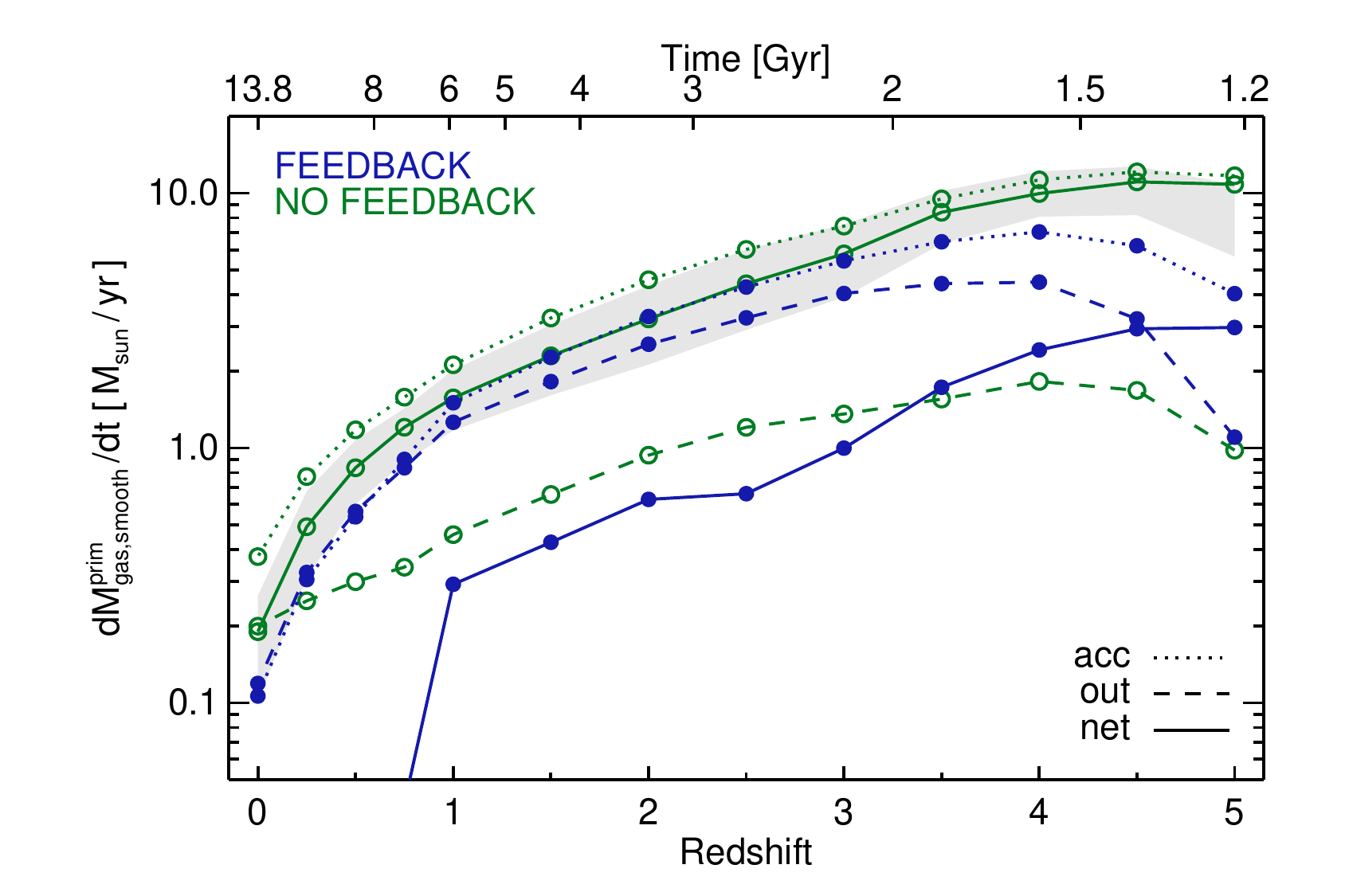}}
\caption{ The smooth gas accretion rate of cosmological origin onto central galaxies as a function 
of redshift. At each redshift accretion over a time window of $\Delta t = 250 \,\rm{Myr}$ contributes. 
We include here only haloes in the mass range $11.3 < \log{(M_{\rm halo,tot}/\rm{M}_\odot)} < 11.4$, and consider separately 
the contribution from net inflow, outflow, and raw accretion. The grey band indicates 
the upper and lower quartiles about the median for the noFB net line. 
 \label{fig_rates_vs_redshift}} 
\end{figure}

To understand the declining net accretion rates towards redshift zero, Fig. 
\ref{fig_rates_vs_redshift} shows the separate measurements of net accretion, outflow, and raw inflow, disregarding 
the comparison between \tmax and \tviracc. The ``outflow'' in the no feedback run 
arises primarily from dynamical gas motions, particularly due to galaxy-galaxy interactions. Any induced velocity 
which moves a tracer outside the 0.15\,$r_{\rm vir}$ radial boundary will result in a non-zero outflow rate. This 
situation could arise during a merger or fly-by which generates any of 
(i) a tidal gas tail, 
(ii) a re-distribution of gas away from the halo centre, 
(iii) turbulent gas motions, 
(iv) or the merging companion itself passing in and subsequently out of this radius.
In some sense this level of outflow represents an unrelated contribution (or noise floor) to the measurement of true, 
feedback induced outflows. In the case of smooth accretion, we see that this contribution is at least a factor of a few 
below the accretion level in the no feedback case. In the feedback run, however, large outflow rates balance about 
half of the inflow, this factor increasing towards redshift zero. The result is the suppression of the net smooth 
accretion rate at all redshifts already noted from the previous figure. Interestingly, the raw inflow rates also differ 
between the two runs, being suppressed in the FB case by a factor of roughly two, again at all redshifts. We find these 
same trends in the other accretion modes, as well as in the total accretion, disregarding mode. They also hold for 
all halo masses down to $\sim\!10^{10.5}$\,\msun, below which the noFB simulation begins to show a similar increase in 
``outflow'' as we reach the limit of sufficiently-resolved galaxies. 

\begin{figure}
\centerline{\includegraphics[angle=0,width=3.6in]{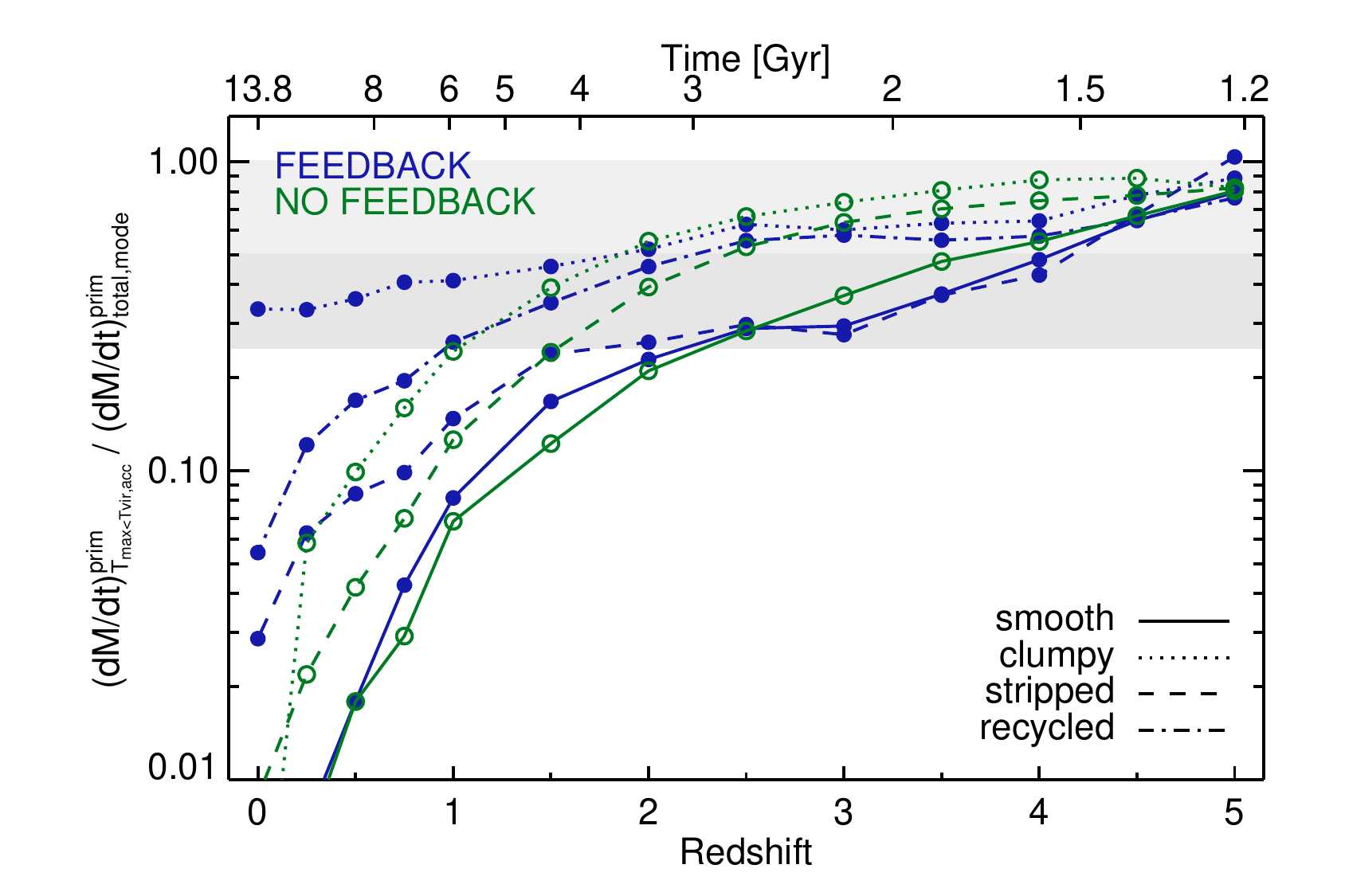}}
\caption{ Ratio of accretion with $T_{\rm max} < T_{\rm vir,acc}$ to the total primordial rate, regardless of $T_{\rm max}$.
We include accreted gas of cosmological origin acquired onto central galaxies over a time window of 
$\Delta t = 250 \,\rm{Myr}$ as a function of redshift. The total inflow is subdivided into smooth, clumpy, stripped, 
and recycled components, where only gas making a net contribution to the accretion rates is included. Grey bands 
indicate 25\%, 50\%, and 100\% levels.
 \label{fig_coldfrac_vs_redshift}} 
\end{figure}

We first consider the ratio of the net rate of the \tmax $<$ \tviracc component to the total primordial net rate, 
independent of past maximum temperature, $f_{<} / f_{\rm tot}$.
This ratio declines monotonically towards redshift zero for all halo masses, such that its maximum value always occurs 
at the highest redshift for which a given halo mass is present in the simulation volume. We explicitly show this ratio 
for each mode in Fig. \ref{fig_coldfrac_vs_redshift}. Contrasting smooth accretion between the two runs at any 
given redshift we find that $f_{<} / f_{\rm tot}$ is nearly identical. The simulation with feedback has smaller  
values at high redshift ($z\!>\!5$), and larger values at low redshift ($z\!<\!2$), the crossover occurring between 
$2\!<\!z\!<\!3$ at all halo masses.
Therefore, while feedback in these systems strongly suppresses the rate of accretion of smooth primordial gas, it does 
not directly affect the temperature history of this material. Although gas which flows in cold may undergo additional 
heating as a result of feedback, this heating is not significant enough to generally increase \tmax above \tviracc. 
Furthermore, gas which flows in hot may suffer a lower \tmax value due to metal-line enhancement of the cooling rates, 
but not at a level to suppress it below \tviracc. The first point implies that outflowing winds do not disrupt inflowing 
streams, at least not enough to induce mixing with the surrounding hot halo gas. Nor does the wind material itself incorporate 
into the streams in a way which modifies their temperature, although here our choice of low thermal energy for stellar 
winds may be partly responsible and potentially masking such an effect. In the following section we consider whether the 
presence of outflows increases the time required for gas to inflow from the virial radius to the galaxy. Here we conclude 
that the temperature history of gas acquired by central galaxies in a smooth mode from the intergalactic medium is largely 
unmodified by our fiducial feedback model.

\begin{figure*}
\centerline{\includegraphics[angle=0,width=7.2in]{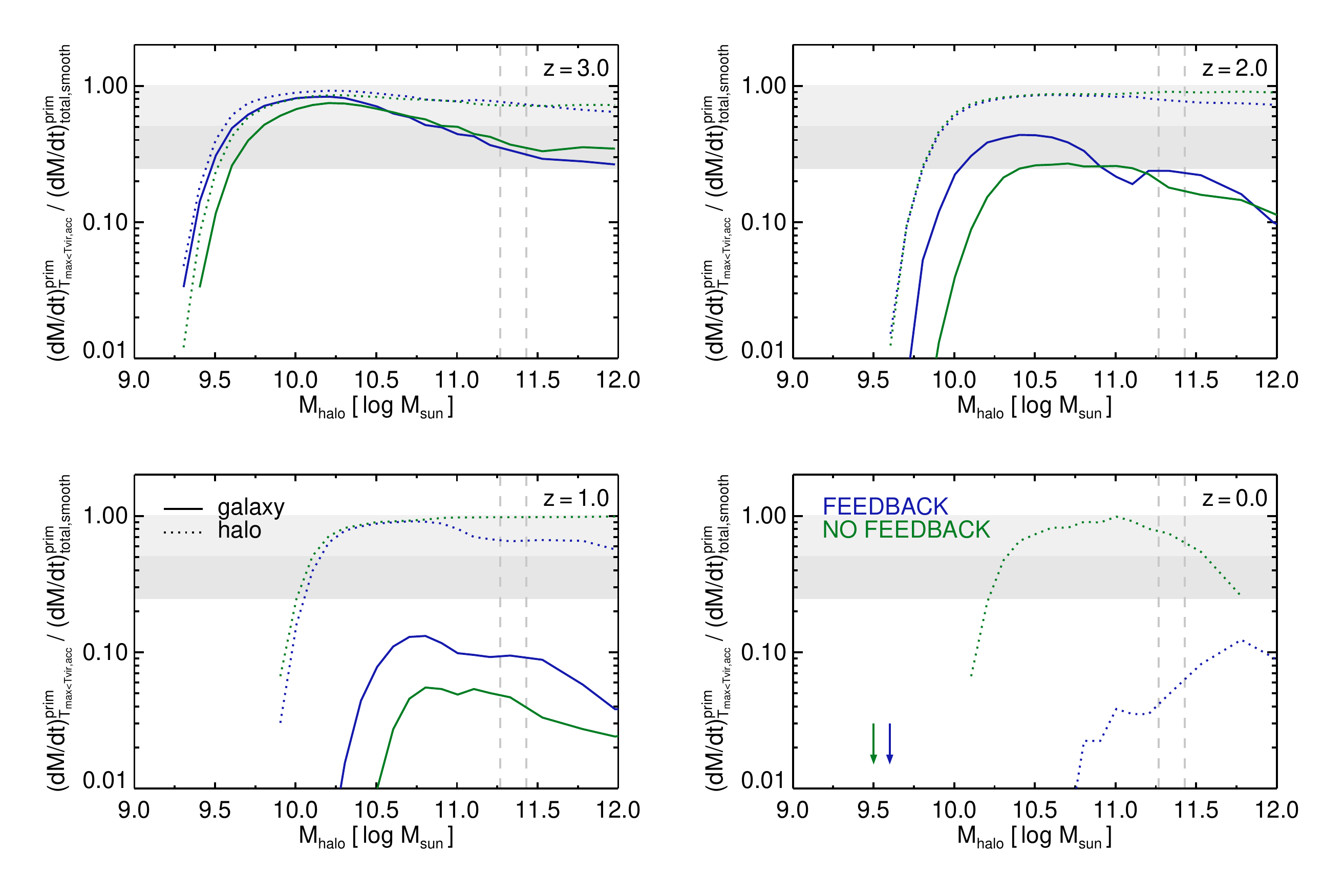}}
\caption{ Ratio of accretion with $T_{\rm max} < T_{\rm vir,acc}$ to the total primordial rate as a function of halo mass.
We include smoothly accreted gas of cosmological origin acquired onto central (solid) galaxies and their parent haloes 
(dotted) over a time window of $\Delta t = 250 \,\rm{Myr}$ ending at four specific redshifts ($z \in \{0,1,2,3\}$).
Only gas making a net contribution to the accretion rates is included. The fractional contribution of the colder 
component onto galaxies evolves downward with redshift largely independent of halo mass, and similarly between the two 
runs. At low redshift, $z\!<\!1$, the FB run suppresses this fraction for gas accretion onto haloes as a result of 
significant heating prior to crossing the virial radius. Horizontal grey bands indicate the 25\%-50\%, and 50\%-100\% 
levels, and dotted vertical lines indicate the mass range considered throughout.
 \label{fig_coldfrac_4redshifts}} 
\end{figure*}

Significantly different behaviour is seen for baryonic mass which accretes as part of a resolved substructure as it 
crosses the virial radius. In both the FB and noFB runs, $f_{<} / f_{\rm tot} \ga 0.5$ for this clumpy material for all 
redshifts $z\!>\!2$. This ratio drops to nearly zero towards $z\!=\!0$ in the noFB run, indicating that material 
incorporated into galaxies which entered the halo bound to a satellite nevertheless experiences heating comparable 
to smoothly accreted gas. In contrast, this ratio only drops to $\simeq$\,0.4 by $z\!=\!1$ and $\simeq$\,0.35 by $z\!=\!0$ 
in the FB run. We propose that metal enrichment in satellites prior to accretion into the MPB enables more efficient 
radiative cooling at later times, preventing a sizeable fraction of this gas from reaching a maximum past temperature 
as high as its counterpart in the noFB run. This would require a gradual heating process, such as the mixing and 
stripping of satellite gas during its orbit through a larger hot halo, processes which remain largely unexplored in the 
present work.

A similar though less prominent effect is also evident for stripped material. Recycled material -- which was part of a 
stellar feedback driven wind at some previous time, but could otherwise be part of any accretion mode -- has a redshift 
evolution comparable to that of stripped gas, supporting the idea that one or more epochs of metal enrichment in the 
ISM of an external galaxy, prior to accretion, can suppress \tmax.

\subsection{Gas acquisition by the halo}

It is notable that for material acquired by the halo itself this same ratio $f_{<} / f_{\rm tot}$ does differ between 
the simulations. In particular, we consider the net flux of tracers crossing 1.0\,$r_{\rm vir}$ for the first time over 
the past 250 Myr, without necessarily also crossing 0.15\,$r_{\rm vir}$. For z\,$<$\,1, in the same halo mass range of 
$11.3 < \log{(M_{\rm halo,tot}/\rm{M}_\odot)} < 11.4$, we find $f_{<} / f_{\rm tot} \simeq 0.9$ in the run without 
feedback versus $f_{<} / f_{\rm tot} \simeq 0.6$ in the run including feedback. We expect this fraction to be large in 
both cases, since the temperature distribution of material near the virial radius is dominated by gas at temperatures 
lower than the virial temperature. Specifically, the median radial temperature profile of the hot halo gas at $z\!=\!0$, 
$T(r)/T_{\rm vir}$, decreases with radius to a minimum of $\simeq$ 0.5 at $r/r_{\rm vir}$\,=\,1. We then expect that 
for most gas the ratio \tmax / \tviracc will reach a maximum of $\sim$\,0.5 in both runs. On the other hand, the exact 
fraction of gas satisfying \tmax < \tviracc could be lower than unity if the gas temperature distribution at the virial 
radius extended into a high temperature tail.

This is not true in the FB case, implying that accreting gas has already evolved following a different 
temperature history prior to interaction with the halo. As a check, we make a simple measurement of the temperature 
of gas in the intergalactic medium, defined as all gas cells outside of all friends-of-friends groups. Between the FB 
and noFB runs, the IGM temperature distribution is similar at high redshift ($z\!>\!3$) and deviates strongly by 
$z\!=\!0$. In particular, while the mean IGM temperature in the noFB case remains essentially constant between 
$0\!<\!z\!<\!2$ at $T_{\rm IGM} \simeq 10^{4.2}$\,K, in the FB run it increases from that value at $z\!=\!2$ to 
$T_{\rm IGM} \simeq 10^{5.3}$\,K by $z\!=\!0$. The primary cause is our black hole feedback model operating in the radio-mode, 
which begins to alter the global IGM temperature statistics in the simulated volume as haloes of sufficiently high mass 
begin forming at $z\!\simeq\!3$. In a larger simulation volume this would take place at higher redshift, and with a less 
homogeneous effect on the global box. We caution, however, that the gas content of massive systems is too low in 
our simulations \citep{genel14}, indicating that although the radio-mode model is efficient at moderating stellar mass 
growth in these haloes, its side effects -- including the strong influence on IGM temperature -- likely imply that the 
details of this model require modification.

\begin{figure*}
\centerline{\includegraphics[angle=0,width=7.2in]{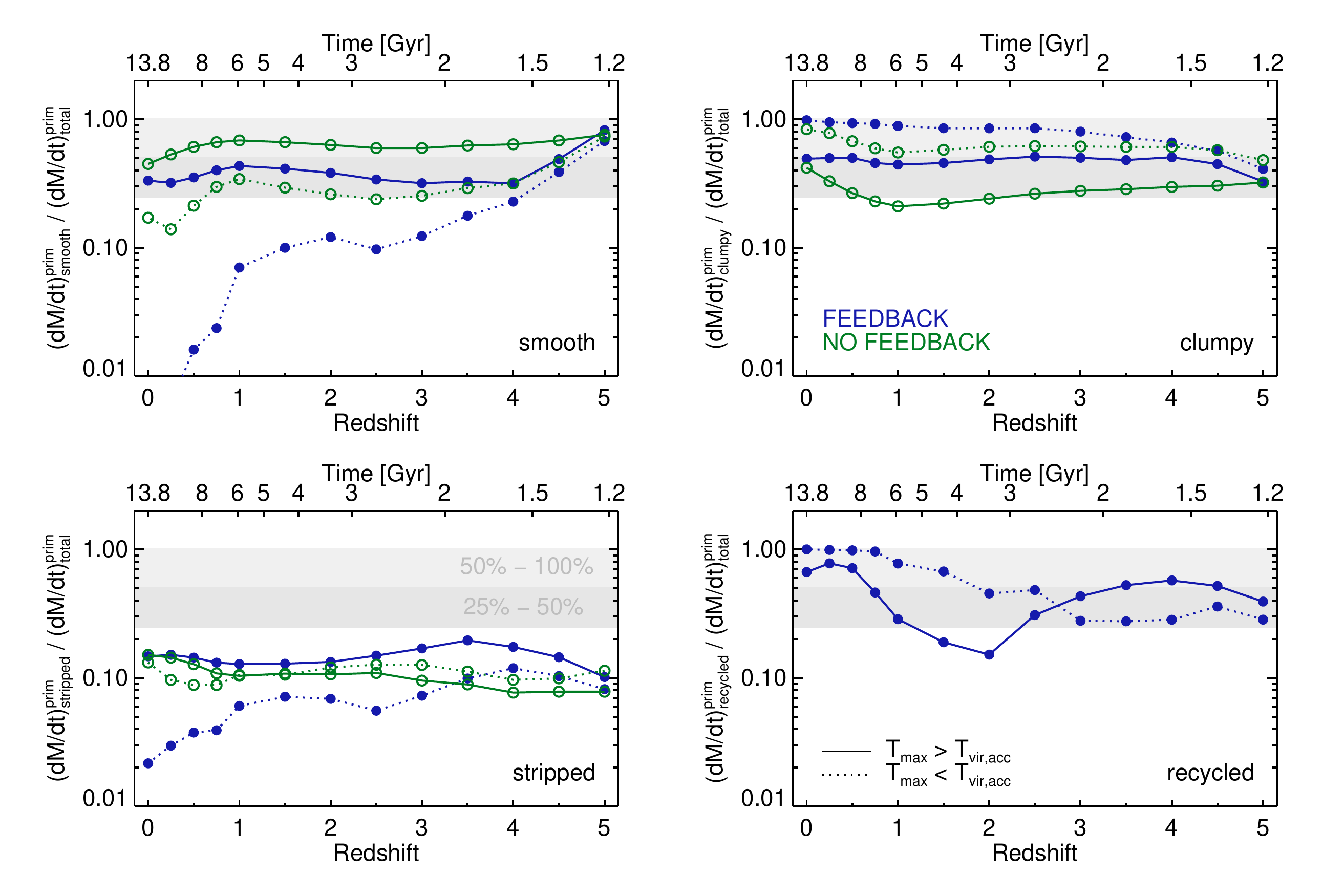}}
\caption{ The fractional contribution of each accretion mode to the total net accretion rate of gas from cosmological 
origin onto central galaxies over a time window of $\Delta t = 250 \,\rm{Myr}$, as a function of redshift. We separate 
the contribution based on the comparison of the maximum past temperature \tmax of each Monte Carlo tracer to the 
virial temperature of future parent halo, at the time of accretion. The grey band indicates the $25\%$ to $50\%$ range. 
The four modes are smooth, clumpy, stripped, and recycled. The feedback run decreases the fractional importance of 
smooth accretion, regardless of temperature history and for all redshifts, particularly for cold material at late times.
 \label{fig_ratefracs_vs_redshift}} 
\end{figure*}

In Fig. \ref{fig_coldfrac_4redshifts} we show how the fraction of gas with \tmax $<$ \tviracc smoothly accreted onto 
haloes develops this difference between the FB and noFB cases as a function of halo mass, for four redshifts. The 
fraction of \tmax $<$ \tviracc material accreted onto haloes agrees at $z\!=\!2$ when comparing the FB and noFB runs. 
However, for sufficiently massive haloes by $z\!=\!1$ this fraction is suppressed by a factor of two in the FB run, and 
drops to 10\% or less by $z\!=\!0$. It is notable that this same difference is \textit{not} seen for accretion onto 
galaxies at low redshift. In this case, for the runs both with and without feedback, the fraction of gas with 
\tmax $<$ \tviracc drops to zero at late time. This indicates that, despite the thermal history differences due to the 
IGM temperature, further heating proceeds within the halo before accretion onto the galaxy. The relative importance 
between a single strong virial shock, a series of smaller shocks within the halo, and adiabatic compression in this 
gas heating process remains an interesting question for future work.

\subsection{Relative importance of different accretion modes}

Having so far focused on smooth accretion, we now consider the importance of the different accretion modes. In 
Fig. \ref{fig_ratefracs_vs_redshift} we show the fractional contribution of each mode to the total net accretion rate 
as a function of redshift. Each of the four modes is split into a separate panel. Each mode is further divided 
based on the comparison between \tmax and \tviracc. Most clearly, the contribution of smooth and clumpy accretion modes moves in 
opposite directions between the two runs. The inclusion of feedback physics suppresses the total relative contribution 
of smooth accretion by a factor of $\sim$\,2. This is true at essentially all redshifts, and for all gas 
regardless of temperature history. However, the contribution of the smooth component to \tmax $<$ \tviracc gas towards 
$z\!=\!0$ is especially modified, being reduced to a negligible amount by the present day. We note that this conclusion 
holds without contradiction given our earlier finding from Figure \ref{fig_netrate_vs_redshift} that the contribution of 
\tmax $<$ \tviracc gas to the smooth component -- that is, the ``cold fraction'' of smooth accretion -- is largely 
unchanged.

The fractional importance of stripped material is similar between the two runs. By definition the hot and cold 
component lines both sum independently to one (excluding recycled, which is an additive attribute). Consequently, 
the contribution of resolved substructures increases to balance the decrease of the smooth mode. In terms of primordial 
accretion contributing to the net growth of galaxies, for most of cosmic time, $z\!<\!3$, more than 80\% (less than 40\%) of 
material with a cold (hot) temperature history is acquired in the clumpy mode. This exact fraction will be sensitive to 
the definition of substructure -- for instance, we include the entirety of a satellite halo, whereas \cite{keres09} 
includes only the actual satellite galaxy/ISM material. Finally, tracers which have at some 
point in their past resided in the wind phase make up a significant fraction of the total net rate, particularly at 
late times -- redshift zero growth is dominated by this recycled component. However, the fact that we are here only 
considering gas which has entered the MPB for the first time implies that this is largely recycling in satellite systems 
prior to incorporation into the central galaxy.


\section{The State of Gas in the Halo} \label{sTwo}

\begin{figure*}
\centerline{\includegraphics[angle=0,width=6.8in]{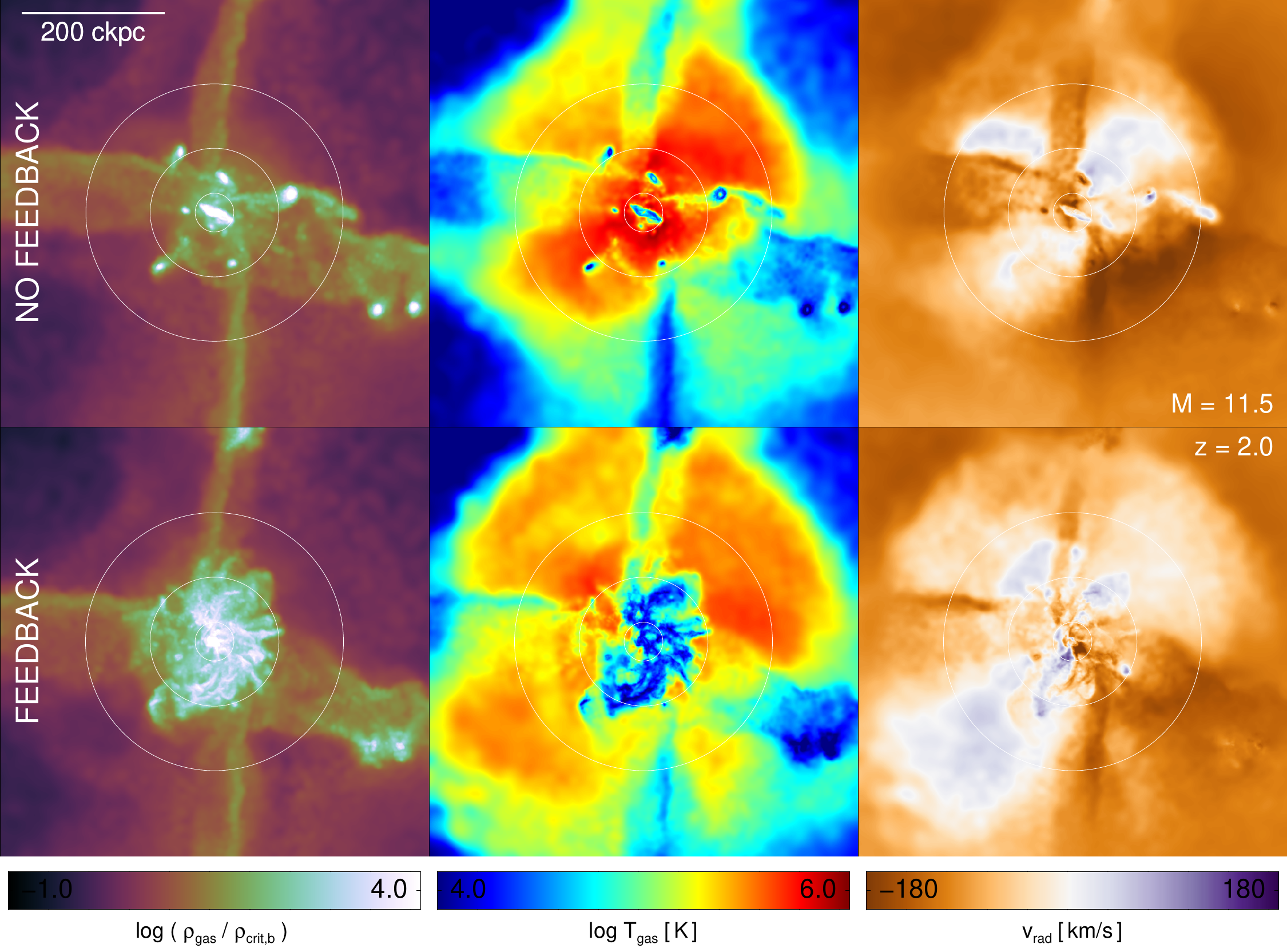}}
\caption{ Comparison of a single halo with a mass of $10^{11.5}$\,\msun\, at $z\!=\!2$ between the FB and noFB runs. Shown 
in projection are gas density (left), mass-weighted temperature (middle), and mass-weighted radial velocity (right). 
The three white circles are the same in all panels and denote $[1.0,0.5,0.15]$ $r_{\rm vir}$. Negative $v_{\rm rad}$ 
(brown) is infall, whereas positive (purple) $v_{\rm rad}$ is outflow. Introducing the feedback model pushes the 
hot halo gas to larger radii, while increasing the fraction of the virial sphere covered by outflow. Gas streams 
inflowing across the virial radius are largely unaffected, in terms of their temperature, density, or radial velocity. 
The virial radius of this halo is $180$\,kpc comoving, and the virial temperature is $\simeq$\,10$^6$\,K.
 \label{fig_image_halocomp}} 
\end{figure*}

Here we consider the instantaneous properties of gas flows in the halo regime, regardless of its past or future 
history. To begin, Fig. \ref{fig_image_halocomp} shows a prototypical, single halo in the mass range under consideration, 
matched between the two runs, at $z\!=\!2$. An orthographic projection with extent in all dimensions equal to 3.5 times the 
virial radius (indicated with the largest white circle) shows gas density, mass-weighted gas temperature, and 
mass-weighted gas radial velocity, where negative denotes inflow. The run without feedback is shown on the top row, 
while the run with feedback is shown on the bottom row.

The temperature projection reveals that the energy injection from feedback arising in the central galaxy pushes the hot 
halo gas to slightly larger radii. The existence of cold and metal enriched wind material in our model suppresses the peak 
temperature in the inner halo, both directly and due to enhanced cooling. The stellar feedback driven 
winds clearly populate the inner halo ($r/r_{\rm vir} < 0.5$) with a large mass of cold gas with high covering fraction, 
substantially altering the temperature and velocity structure of gas at these small radii. Further from the galaxy, we 
see that more of the halo volume -- and halo gas mass -- is occupied by material with outward radial velocity. However, 
at the virial radius, gas inflow appears largely unaffected by the introduction of our fiducial feedback model. While in 
mass-projection the streams crossing $r_{\rm vir}$ appear somewhat more collimated, slower, and warmer, this is 
predominantly a side effect of this particular visualization. If we instead inspect spherical slices of gas properties 
at the virial radius, we find that there is e.g. no notable difference in the temperature distribution of inflowing material.

\begin{figure}
\centerline{\includegraphics[angle=0,width=3.2in]{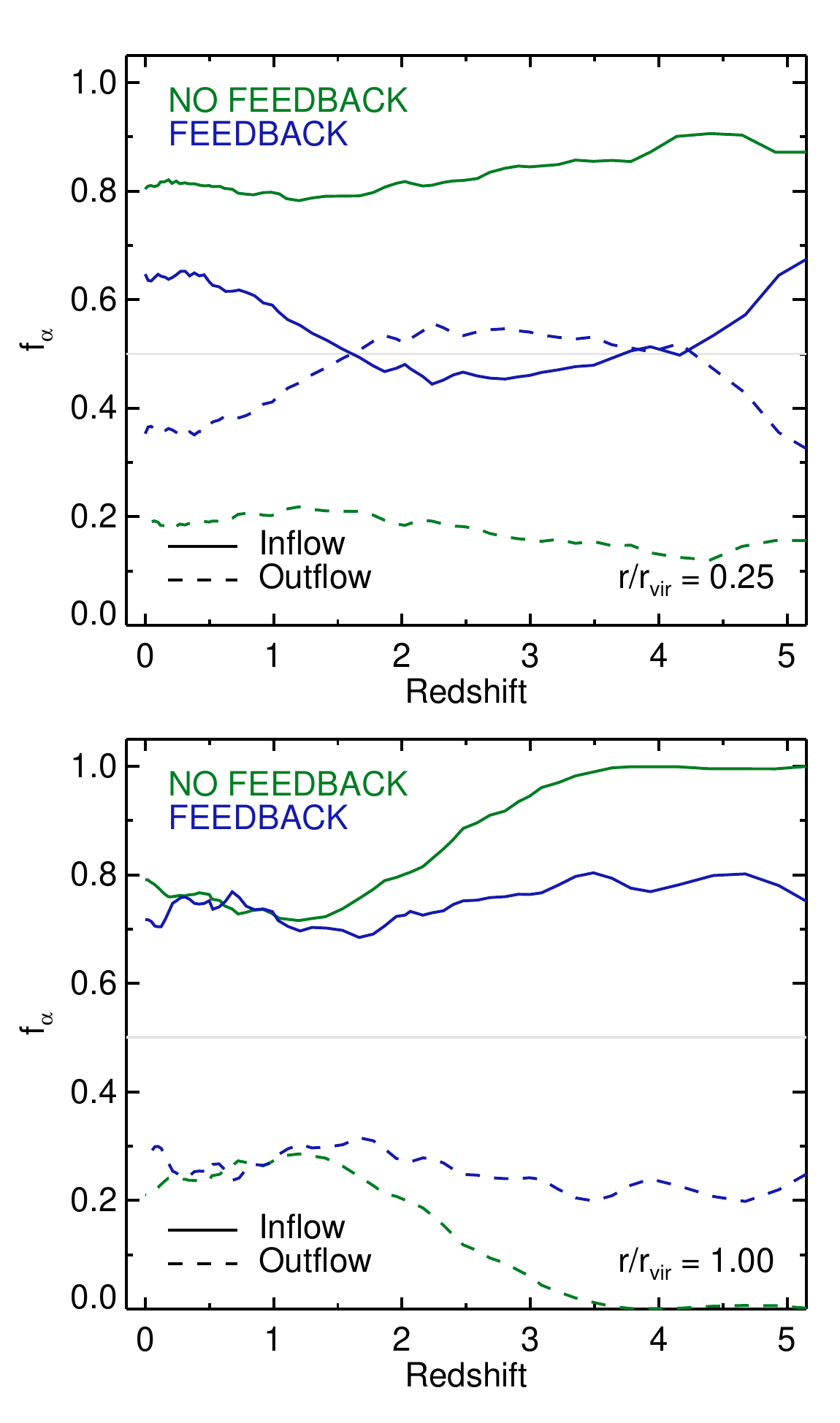}}
\caption{ The spherical covering fraction of the quarter virial sphere (top panel) and full virial sphere (bottom panel) 
of inflow and outflow. Each is calculated as the fraction of the surface covered by gas of either positive or negative 
radial velocity which exceeds a relative threshold of 10\% of the circular velocity of an NFW halo at that radius. 
Gravitationally bound substructures have first been removed. The halo mass range of 
$11.3 < \log{(M_{\rm halo,tot}/\rm{M}_\odot)} < 11.4$ is included. Near the galaxy, the presence of feedback increases 
the covering fraction of outflow above half, while at the virial radius this same behaviour is seen to a lesser degree 
and only at high redshift.
 \label{fig_fracinout}} 
\end{figure}

To be more quantitative, we construct spherical slices for all haloes at all analysis redshifts at a number of radii. 
We use a mass-weighted tophat kernel to interpolate gas quantities onto equal area pixels using the {\small Healpix} 
scheme \citep{gorski05}. In Fig. \ref{fig_fracinout} we calculate a spherical covering fraction of inflow or outflow 
as the fraction of pixels on this sphere with radial velocity above some threshold with the appropriate sign. As a 
threshold we take 10\% of $v_{\rm circ}$(r) of an NFW halo of equal mass. Our results are insensitive to this choice 
provided it is small enough --  we also considered a constant threshold of $\sim$\,20\,km\,s$^{-1}$, and a threshold of 
zero. The top panel shows the behaviour across the quarter virial sphere, while the bottom panel shows the same across 
the full virial sphere. In all cases we exclude all gas bound to resolved substructures prior to this calculation.

At 0.25\,$r_{\rm vir}$, the impact of feedback is to substantially reduce the spherical covering fraction of inflow 
from 80\%\,-\,90\% down to 50\%\,-\,60\%, while outflow correspondingly covers more than half of the surface at $z\!\sim\!2\!-\!3$, up 
from $\simeq$\,20\%. At the virial radius, both runs agree that $\ge$70\% of the virial sphere is covered by inflow, 
regardless of redshift. In the FB case, somewhat more material has outward radial velocity at high redshifts, but for 
this halo mass regime the difference is small and disappears towards $z\!=\!0$. In agreement with the single halo 
shown previously in Fig. \ref{fig_image_halocomp}, we see that at $z\!=\!2$ feedback introduces a negligible change to the 
spherical covering fractions of both inflow and outflow at the virial radius.

\begin{figure}
\centerline{\includegraphics[angle=0,width=3.2in]{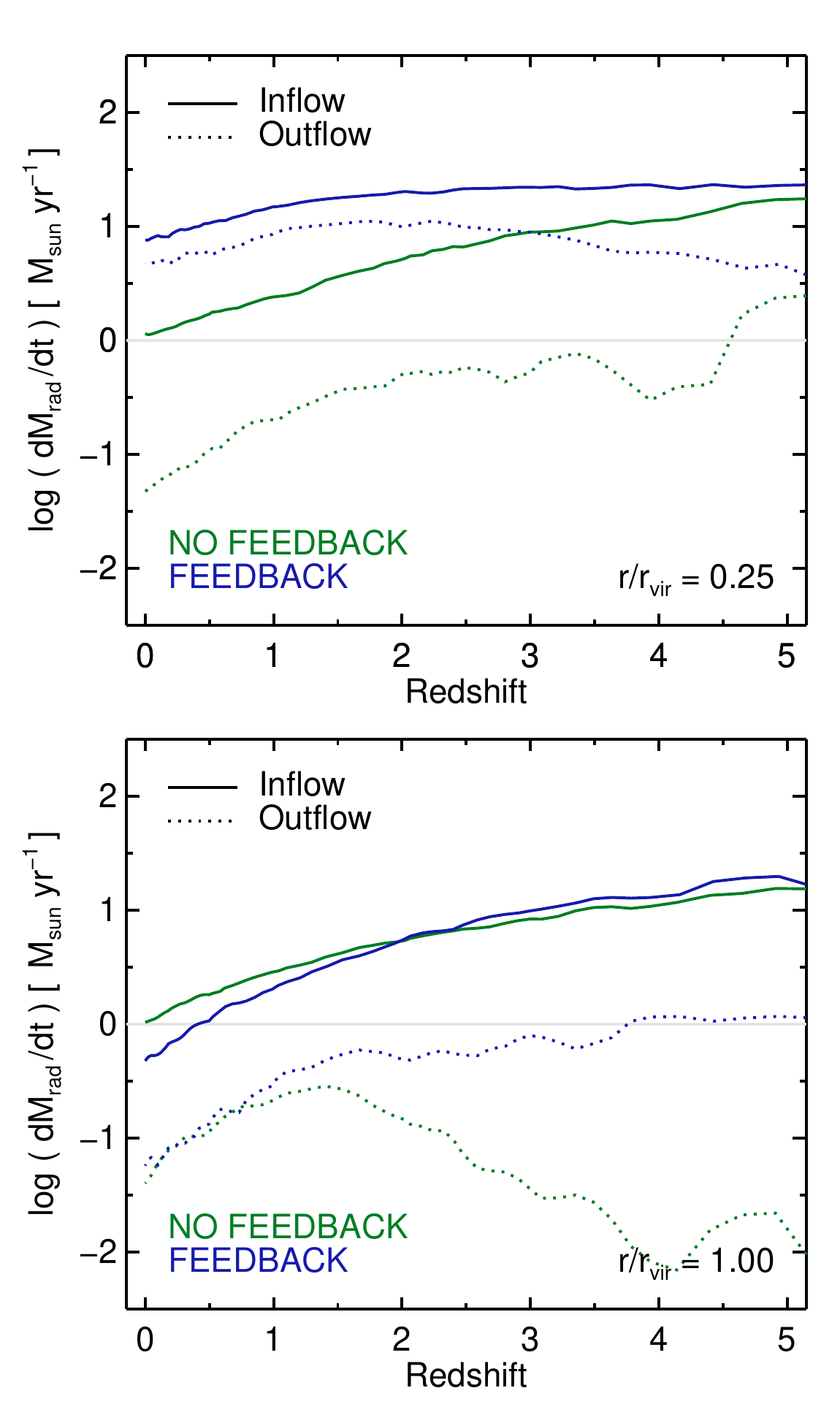}}
\caption{ The radial mass flux rates across the quarter virial sphere (top panel) and full virial sphere (bottom panel), 
calculated separately for inflow and outflow. As in the previous figure, gravitationally bound substructures have first 
been removed, and the halo mass range of $11.3 < \log{(M_{\rm halo,tot}/\rm{M}_\odot)} < 11.4$ is considered. While the rate of 
inflow across the virial sphere is unchanged, the rates of both inflow and outflow closer to the central galaxy are 
both significantly increased, implying recycling occurring across the 0.25\,$r_{\rm vir}$ boundary.
 \label{fig_rateinout}} 
\end{figure}

In Fig. \ref{fig_rateinout} we calculate the radial mass flux rate of gas across the same two radii, for both inflow and 
outflow, with substructures removed. Since inflow and outflow occupy disjoint portions of each sphere, these rates are 
in principle independent, and do not necessarily correlate with the respective spherical covering fractions. At the 
virial radius (bottom panel) we see that the rate of gas inflow is unchanged, while feedback boosts outflow rates at 
high redshift substantially, although to magnitudes which are still small relative to inflow. This outflow across the 
virial sphere transfers mass, and therefore metals, from within virialised haloes out into the intergalactic medium. 
Notably, the presence of feedback does not lower the instantaneous inflow rates, which would indicate a direct impact 
on the accretion of intergalactic material onto the halo. Nor does it increase the inflow rates, which would indicate 
large-scale recycling motion across the host halo virial radius.
At 0.25\,$r_{\rm vir}$ (top panel) feedback has a larger impact, increasing the rates of both inflow and outflow, the 
magnitude of the difference growing with time. Given that the fraction of this surface covered by inflow is actually 
smaller in the FB run, this is a clear signature of significant gas recycling motion across this boundary.


\section{Timescale of Gas Accretion} \label{sThree}

\begin{figure}
\centerline{\includegraphics[angle=0,width=3.5in]{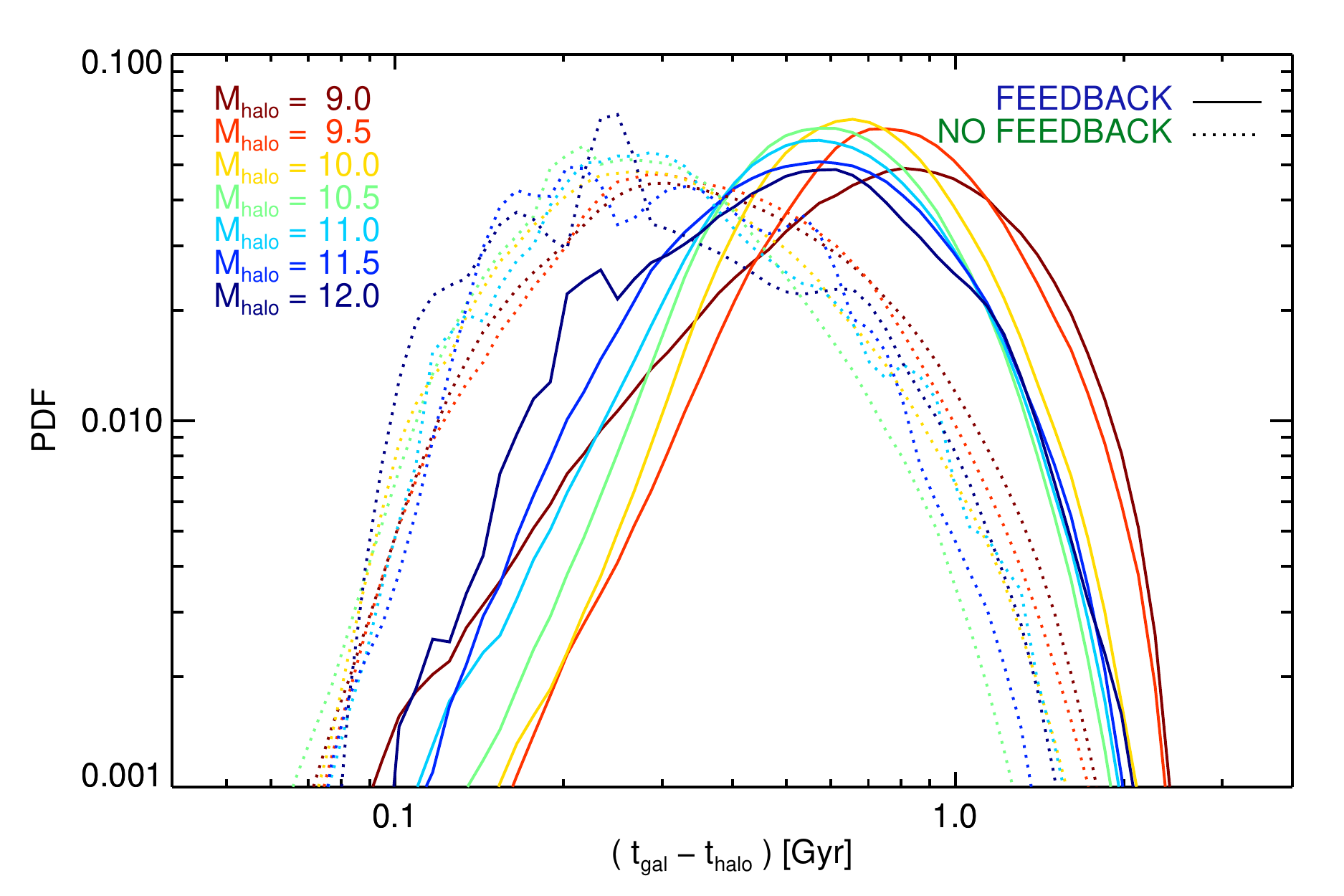}}
\caption{ Distribution of time difference between the first 1.0\,$r_{\rm vir}$ and 0.15\,$r_{\rm vir}$ crossings (the 
``halo transit time'' for each tracer during its first accretion into the halo) for smooth accretion only. Includes all 
tracers within haloes which have recorded these two crossings by $z\!=\!2$. Separated by halo mass (colored lines) and 
for the no feedback (dotted) and feedback (solid) runs. The FB run shifts each distribution, regardless of halo mass, 
to longer times by a constant factor of $\sim$\,2\,-\,3. In addition, for both simulations, the distribution of transit 
times is nearly independent of host halo mass.
 \label{fig_acctimedelta}} 
\end{figure}

In the previous section we saw that the morphology of inflow at the virial radius is largely unchanged. Here, we 
consider whether the presence of outflows increases the time required for gas to inflow (or ``transit'') from the 
virial radius to the galaxy. We measure this quantity as the time difference between the first virial radius crossing 
$t_{\rm halo}$  and the first incorporation into the galaxy $t_{\rm gal}$. In Fig. \ref{fig_acctimedelta} we show the 
distribution of this time difference for all tracers within haloes at $z\!=\!2$ which have previously recorded these two 
crossing times, split into halo mass bins from 10$^9$\,\msun\,to 10$^{12}$\,\msun. We include only smooth accretion.

First, in both runs and for all halo masses we find a broad, uni-modal distribution. That is, there is no obvious 
evidence for multiple channels of accretion having different halo crossing timescales. Comparing the FB (solid) and 
noFB (dotted) runs, we find that feedback introduces a significant delay in the halo transit time of smooth mode 
material. Each distribution shifts to longer times by a factor of $\sim$\,2\,-\,3, and this factor is largely independent 
of halo mass, at least above $10^{10}$\,\msun, where systems are well resolved. Specifically, gas smoothly accreted by 
$z\!=\!2$ takes on average $\simeq$\,250\,Myr to cross from the virial radius to the galaxy in the noFB run, and 
$\simeq$\,700\,Myr in the FB run.

Of even more interest, we see that in both simulations, this halo transit time is again largely independent of halo 
mass. This implies that this time difference may be related more to the dynamical time and \textit{not} to the cooling 
time, since the latter scales strongly with halo mass. Over the halo mass range of $10^{10}$\,\msun\,to $10^{12}$\,\msun, 
then, we see no evidence for a transition point above and below which the process of gas accretion through the halo 
occurs in a fundamentally different way, at least insofar as is captured by our measured ``halo transit time''. We 
return to this point and its implication for the idea of a critical halo transition mass in the discussion.

\begin{figure}
\centerline{\includegraphics[angle=0,width=3.3in]{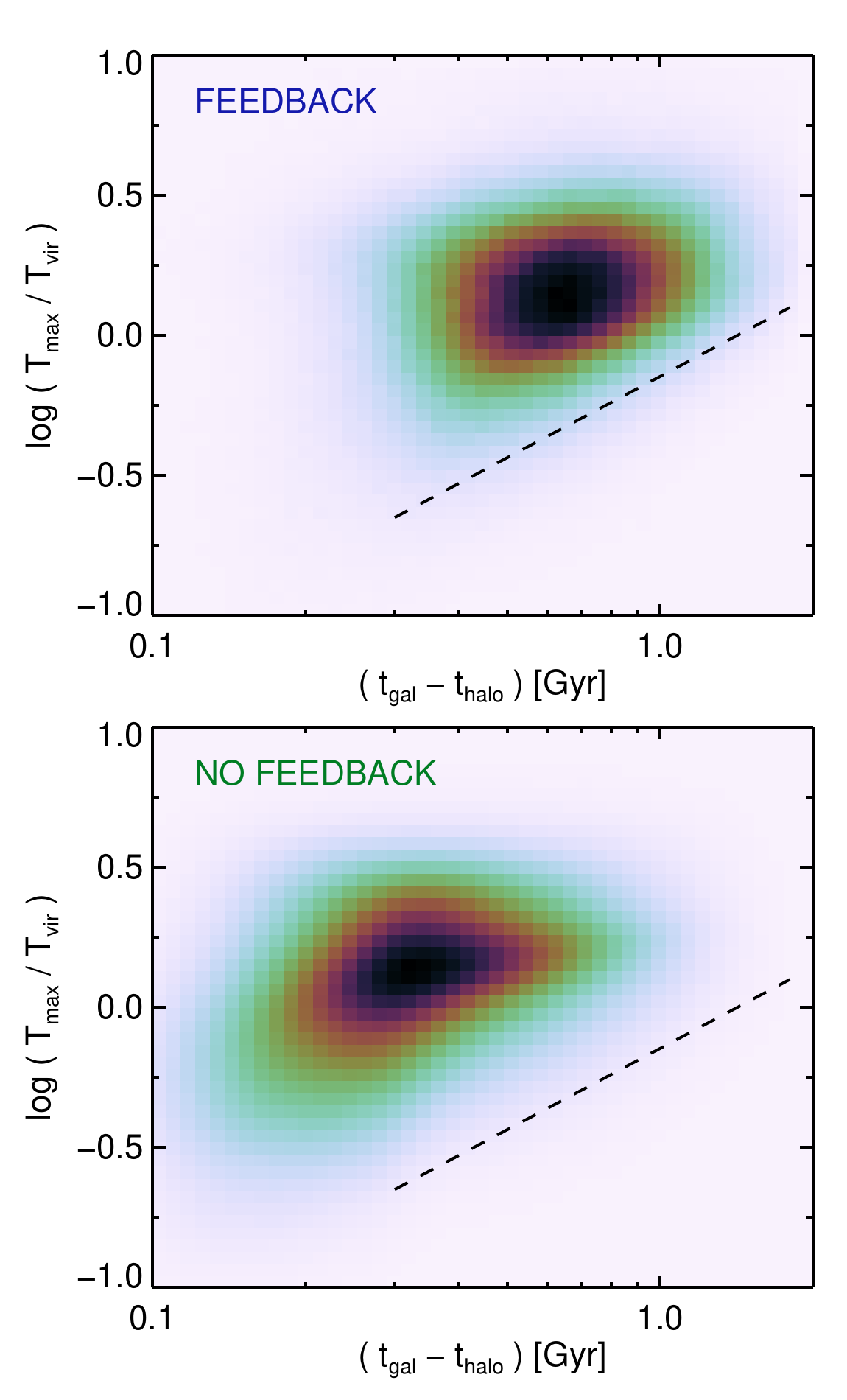}}
\caption{ Correlation between \tmax and ``halo transit time'' ($t_{\rm gal}$-$t_{\rm halo}$), where the colour scale 
indicates the mass distribution of tracers smoothly accreted by $z\!=\!2$ in this plane. While the noFB run (bottom panel) 
may indicate a relation between maximum past temperature and halo transit time, this is less clear in the FB run (top 
panel). In both cases there is a lower envelope, approximately indicated by the dashed line in the top panel, 
indicative of a minimum \tmax which increases with increasing transit time.
 \label{fig_acctimedelta_tmax}} 
\end{figure}

To further explore any possible relationship between this transit time and different accretion mechanisms, 
Fig. \ref{fig_acctimedelta_tmax} shows the relation with the \tmax of each tracer. The question is whether or not gas 
with low \tmax has a shorter halo transit time, while gas with higher \tmax spends longer in the halo, as might be 
naively expected. We see a weak correlation in this direction for the noFB simulation (bottom panel), which is less 
clear after including feedback (top panel). This lack of correlation implies that the mean halo transit times for both 
hot and cold gas are comparable. That is, there is no strong signature of the past thermal state of gas in its 
dynamical history, at least insofar as is measured by the halo transit time. In both simulations, notably, there is 
clearly a lower envelope. This is indicated in the top panel by the dashed black line, which shows the relation

\begin{equation}
\log{\left( \frac{ T_{\rm max} }{ T_{\rm vir,acc} } \right)} = \frac{(t_{\rm gal}-t_{\rm halo})}{2\,\rm{Gyr}} - 0.8 .
\end{equation}

We see that gas does not populate the region to the lower-right of this relation -- that is, there is a minimum \tmax 
reached which increases as a function of increasing halo transit time. Note that we have here stacked together haloes 
of all masses, but we see this relation also when we examine small bins in halo mass. This implies that there is at 
least some link between virial heating and the gas dynamics of accretion. We therefore want to understand where the 
maximum temperature of each tracer is reached -- is the \tmax\, event closely related to the virial crossing time, or does 
it occur on average with either a positive or negative relative lag.

\begin{figure}
\centerline{\includegraphics[angle=0,width=3.4in]{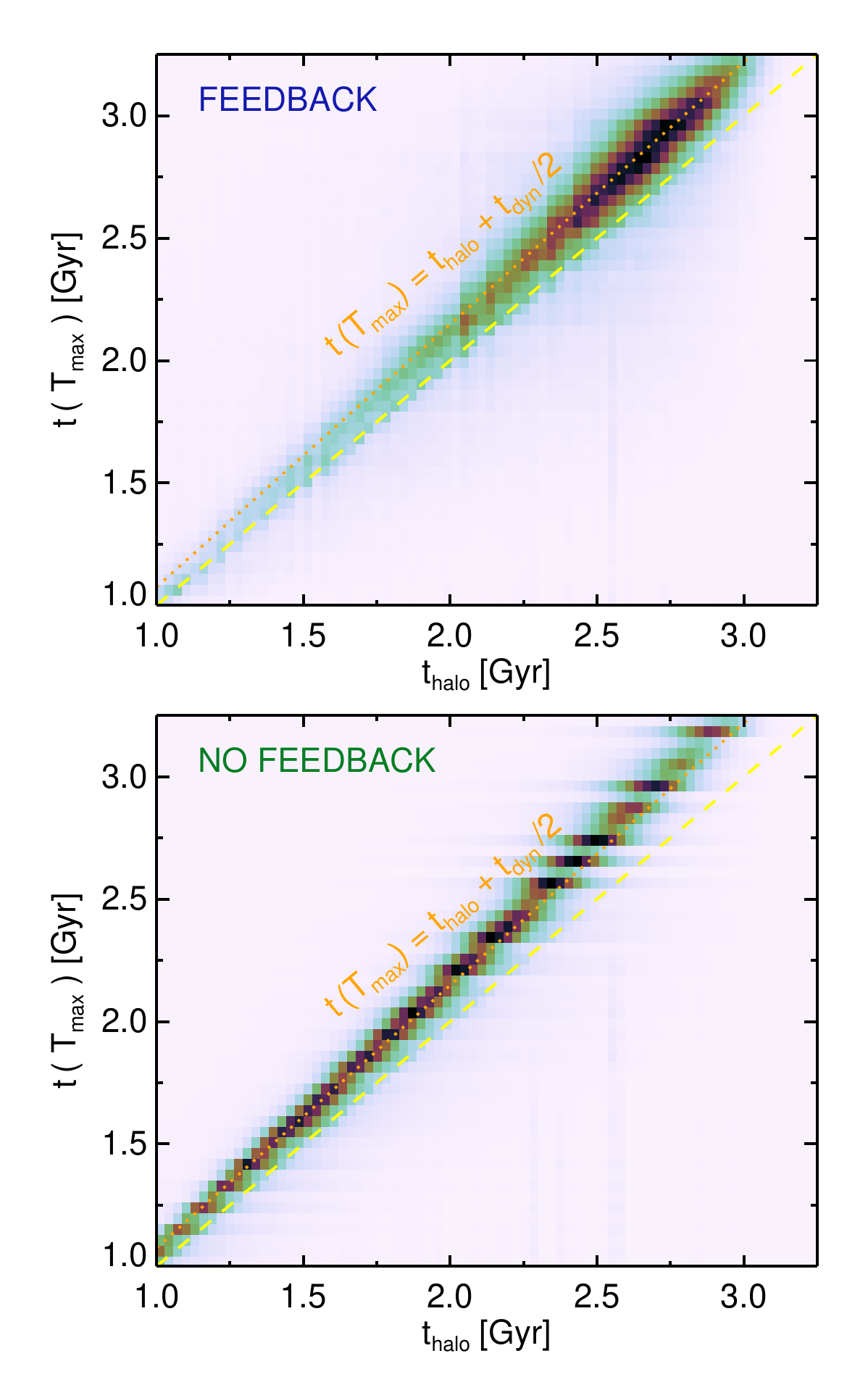}}
\caption{ The tracer mass distribution in the plane of $t$(\,\tmax), the time when the maximum past temperature was 
obtained, and $t_2$, the time of first virial crossing. Both are given in terms of the age of the universe, where 
the yellow dashed line shows the 1-to-1 relation. The orange dotted line is offset by a constant factor, such that 
tracers on this relation would have reached their maximum temperature a time $t_{\rm dyn}/2$ after crossing the 
virial radius. Smooth accretion only, for tracers which have entered haloes by $z\!=\!2$.
 \label{fig_acctime_tmaxtime}} 
\end{figure}

In Fig. \ref{fig_acctime_tmaxtime} we show the correlation between the time of \tmax\, and the time of the first virial 
radius crossing, both in terms of the age of the universe, on a tracer by tracer basis. Only smooth accretion is included. 
The yellow dashed line shows the 1-to-1 line, which would imply that gas heats at $r_{\rm vir}$ to its maximum 
temperature, subsequently cooling in order to join the ISM of the galaxy. The dotted orange line shows a constant 
positive time offset of $t_{\rm dyn}$/2 later, where we take the redshift dependent 

\begin{equation}
t_{\rm dyn} = r_{\rm vir}/v_{\rm circ}(r_{\rm vir}).
\end{equation}

\noindent In both runs, these two lines bound the majority of accreted material, implying that gas reaches its maximum 
temperature shortly \textit{after} first accretion into the halo, and never before. It is clear that the ratio 
$t(T_{\rm max})\,/\,t_{\rm halo}$ increases with time, but since $t_{\rm dyn}$ also evolves with redshift, becoming 
longer at later times, we can see that the characteristic lag time is $\simeq$\,0.5\,$t_{\rm dyn}$, at least for all 
accretion which has occurred by $z\!=\!2$. Comparing the two panels, we conclude that the relation between heating and 
virial crossing for smoothly accreted gas is not strongly affected by the presence of our fiducial feedback model. This 
result holds also for accretion taking place by $z\!=\!1$ and $z\!=\!0$, although the mean lag time increases to 
$\simeq$\,1.0\,$t_{\rm dym}$ for gas entering the halo at late times ($z\!<\!1$).


\section{Discussion} \label{sDiscussion}

\subsection{The contribution of recycled gas} \label{ssArm0}

\begin{figure}
\centerline{\includegraphics[angle=0,width=3.6in]{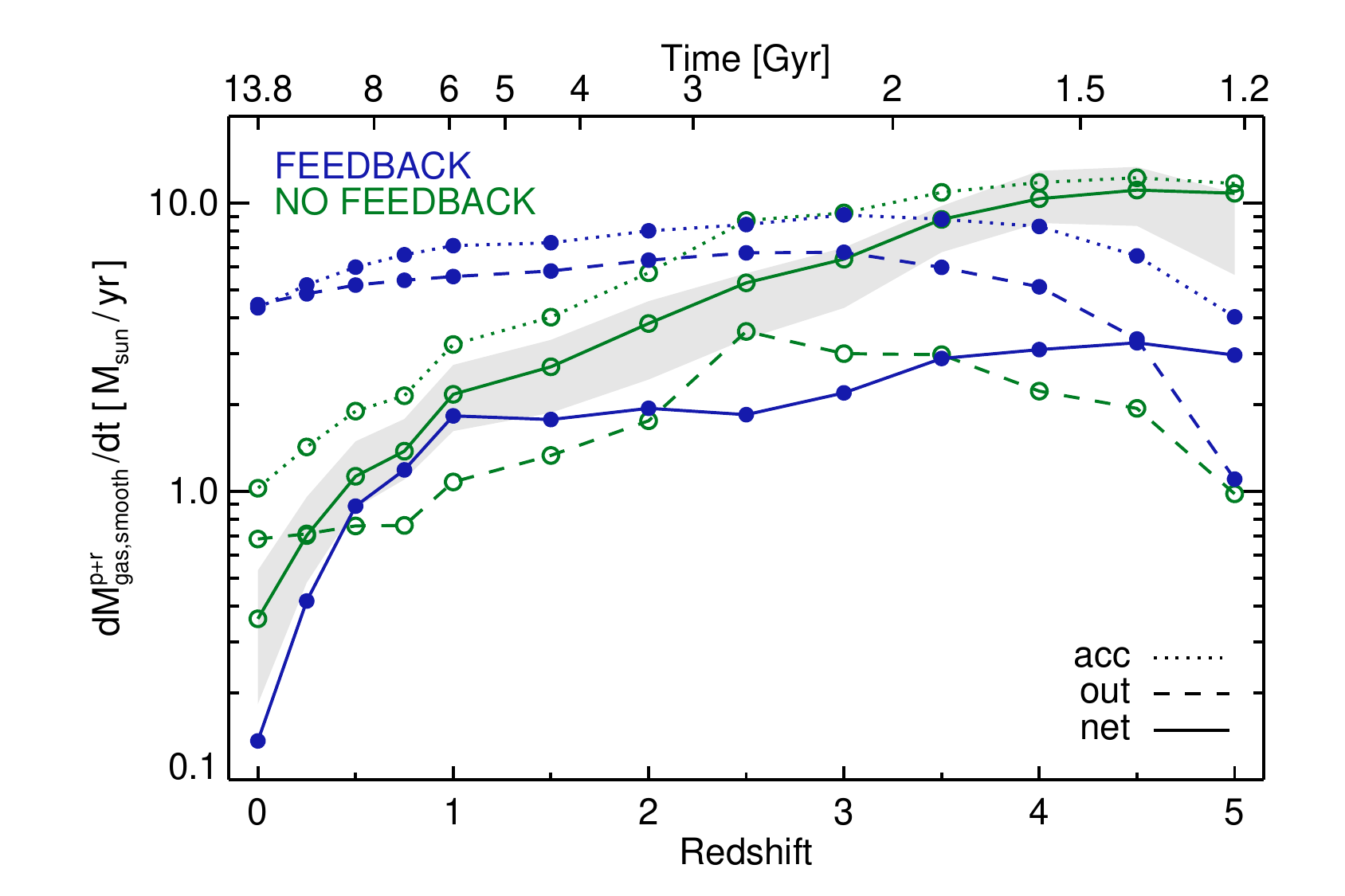}}
\caption{ As in Fig. \ref{fig_rates_vs_redshift} but showing the combined smooth gas accretion rate, including both 
primordial and material previously cycled through the main progenitor branch. We include accretion onto central galaxies 
as a function of redshift. As before, we include only haloes in the mass range $11.3 < \log{(M_{\rm halo,tot}/\rm{M}_\odot)} < 11.4$, 
and consider separately the contribution from net inflow, outflow, and raw accretion.
 \label{fig_rates_vs_redshift2}} 
\end{figure}

We have so far focused exclusively on the accretion of material of primordial, or cosmological, origin.
We can contrast this evolution of $\dot{M}_{\rm gas}^{\rm prim}$ from Section \ref{sOne} with the fundamentally different net 
accretion rates obtained if we count tracers with \textit{most recent} incorporation times, as opposed to \textit{first} 
incorporation times, within the past 250 Myr. We denote this measurement $\dot{M}_{\rm gas}^{\rm p+r}$, where the combined 
contributions from primordial and material previously cycled through the MPB are two disjoint subsets of accretion which 
sum to the total instantaneous accretion rate. As a result, this measurement includes all gas motion across the boundary 
defining the galaxy, regardless of when that gas initially entered the galaxy. Note that by definition, tracers within 
the galaxy with outward crossings are a strict subset of tracers within the galaxy with inward crossings. They represent 
the mass component which cycles out of, and back into, the galaxy during the 250 Myr time interval. 

Fig. \ref{fig_rates_vs_redshift2} revisits the smooth accretion rate as a function of redshift, where the relationship 
between the FB and noFB runs is significantly altered when compared to Fig. \ref{fig_rates_vs_redshift}.
As expected, inflow and outflow rates are higher for both runs at low redshift. The scaling of inflow rates with redshift  
in the noFB run remains similar, with the net accretion rate declining monotonically from $z\!\simeq\!3$ to $z\!=\!0$. 
However, the inclusion of feedback maintains a nearly constant net accretion rate until $z\!=\!1$. The smooth ``p+r'' 
accretion rate at high redshifts, $z\!>\!4$, is similar to the primordial accretion only value, implying that the 
contribution of recycled gas is negligible at early times. However, recycling becomes more important towards low redshift, 
at a level sufficient to balance the decreasing inflow of primordial gas. Between $z\!=\!1$ and $z\!=\!0$ the raw inflow 
rates decrease by factor of $\sim$\,2 in both simulations, pulling the net acquisition of smooth mode material down by a
similar factor.

We can also reconsider the ratio of accretion with $T_{\rm max} < T_{\rm vir,acc}$ to the total rate, as in Fig. 
\ref{fig_coldfrac_vs_redshift}, but including the recycled component (not shown).
In the feedback run, the fraction of material with $T_{\rm max} < T_{\rm vir,acc}$ is larger at all redshifts. 
In particular, at $z\!=\!0$, the ratio to the total is $\simeq$\,0.04, 0.15, 0.25 and 0.4 for smooth, stripped, recycled 
and clumpy modes, respectively. The greater fraction of accretion with a cold temperature history reflects a simple time 
preference -- that recycled gas on average enters the halo earlier than non-recycled gas, combined with a cold fraction 
which increases towards higher redshift.

The other interesting difference we find when including the recycled component in our analysis appears in the 
fractional contribution of each accretion mode to the total net rate, as in Fig. \ref{fig_ratefracs_vs_redshift}.
In this case, the conclusions for hot material with $T_{\rm max} > T_{\rm vir,acc}$ are all qualitatively unchanged, 
whereby feedback still suppresses the relative importance of smoothly accreted gas by a similar amount. However, for 
colder material ($T_{\rm max} < T_{\rm vir,acc}$) the fraction contributed by each of the smooth, clumpy and stripped 
modes converges to similar values for the FB and noFB runs. The reason remains somewhat unclear, and may be coincidental. 
As expected, the recycled mode also becomes more significant, since we now include recycling within the main progenitor 
branch. In this case, the contribution is monotonically increasing towards low redshift, increasing from $z\!=\!5$ to 
$z\!=\!0$ from 0.3 to 1.0 (cold) and from 0.6 to 1.0 (hot). That is, by redshift zero, essentially all gas being accreted 
by galaxies in this mass regime has been previously recycled at least once.

An in-depth consideration of the recycled mode is beyond the scope of this paper, and remains a topic for future work. 
Of particular interest is the recycling timescale -- that is, the amount of time between ejection in a wind and 
re-incorporation into a galaxy. This quantity is particularly important in semi-analytic models, for instance to fix the 
overly early build-up of low mass galaxies \citep{henriques14}. Further, our analysis methodology, which revolves around 
recording only one radial crossing time per direction per radius per tracer, implicitly assumes that gas does not cycle 
through the same radius more than once over the time interval (250 Myr). \cite{opp10} found, for instance, a recycling 
time of $\sim$\,1\,Gyr at $z\!=\!1$ for haloes of this mass. Also of interest is the prevalence of multiple recyclings, 
or, the number of times a baryon mass element has belonged to a wind in its history. Similarly, the contribution of 
satellite recycling and satellite winds in general as a contributor to both the intergalactic medium and the circumgalactic 
medium of larger host haloes. 

However, the process of recycling involves the complex wind physics associated with feedback as well as the difficult 
numerical problem of multiphase gas interaction \citep{agertz13,sarkar14}. More faithful modelling of stellar feedback 
processes and/or higher resolution than is currently available in full cosmological volumes is likely required before 
the process of recycling can be better understood \citep{hop14fire,aumer13,stinson13,fg14,agertz14,cev14}. In more 
massive haloes, treatments of AGN feedback will also need to be improved \citep{wurster13,costa14,bachmann14,li14}, 
while the contribution from other outflow mechanisms such as cosmic rays \citep{booth13,hanasz13,salem14} is potentially 
also critical. Clearly, the baryon cycle in its entirety cannot yet be satisfactorily understood given the current level 
of sophistication of feedback modelling in cosmological simulations.

\subsection{Implications for a critical halo transition mass}

In all semi-analytic models of galaxy formation \citep[e.g.][and more recent efforts]{wf91,croton06,bower06}, a 
foundational theoretical consideration is the cooling of hot halo gas. This essentially governs the growth of the cold 
mass reservoir of the galaxy, and so the subsequent star formation and stellar mass growth. A common approach is to 
let the hot halo gas with $t_{\rm cool} < t_{\rm dyn}$ or $t_{\rm cool} < t_{\rm H}$ cool onto the central galaxy.

Furthermore, they generally implement a critical halo transition mass, below which hot halo cooling proceeds in a 
``rapid'' mode, and above which cooling proceeds in a ``slow'' mode, where the gas fraction eligible to cool drops to 
zero. This difference in the behaviour of quasi-static hot halo gas between more and less massive dark matter haloes 
is motivated largely from the classic, analytical arguments of \cite{silk77,rees77,wr78}.
Analytic arguments supported by one-dimensional, spherically symmetric calculations, cast in terms of whether or not a 
stable virial shock can exist \citep{bd03,db06} have also identified a transition point in halo mass. They find that 
haloes below $\sim$10$^{11}$\,\msun, independent of redshift, cannot provide enough post-shock pressure to support a 
virial shock against losses from radiative cooling.

In Section \ref{sThree} we found that for both simulations, with and without feedback, the time taken by baryons 
to transit from the virial radius to the central galaxy is independent of halo mass \citep[see also][]{brook14}. This 
holds true, at the very least, for haloes with $10^{10}$\,\msun\, $<$ M$_{\rm halo} < 10^{12}$\,\msun. This indicates 
that, in contrast to analytic expectations, we see no strong evidence in the simulations for a sharp halo transition 
mass, above and below which the process of gas accretion proceeds in a fundamentally different manner. It will be 
important to validate this result in the future by extending the dynamic range of this analysis to more massive haloes. 
In particular, good sampling is available up to $10^{14}$\,\msun\,in the $\simeq$\,(100 Mpc)$^3$ volume of Illustris.

The agreement between theoretical cooling models and hydrodynamic simulations has been an occasional subject of 
interest over the past decade \citep{yoshida02,viola08,lu11,monaco14}, where the general conclusion has been that the cooling 
models implemented in SAMs provide an acceptable match to simulations. However, these studies have all used the smoothed 
particle hydrodynamics (SPH) technique, and (as \cite{monaco14} points out) in \cite{nelson13} we found that numerical 
inaccuracies in ``classic'' SPH strongly affect the processes of gas accretion in cosmological simulations, particularly 
failing to accurately capture the cooling properties of hot haloes. The very fact, then, that older SPH simulations 
agree with classic hot halo cooling theory points to a looming disagreement between that theory and modern, cosmological 
simulations. Understanding in what ways these simple theoretical arguments agree with numerical calculations 
\citep[see also][]{dekel13}, and how the picture of hot halo cooling in semi-analytical models can potentially be 
improved upon, remains an open, important question for future work.

\subsection{Comparison to previous studies}

As pointed out in the introduction, reconciling previous conclusions is complicated by a variety of different feedback 
implementations, numerical methods, halo mass regimes, statistical samplings, and interpretations. We compare our 
findings with those works which specifically focused on the impact of feedback on cosmological gas accretion.

In particular, \cite{opp10} considered a few variations of kinetic wind schemes similar to the one presented herein, 
with constant velocities or a momentum-driven scaling, and decomposed star formation rates into three distinct modes: 
hot, cold and wind (recycled). Contrasting recycled versus non-recycled, they concluded that accretion is dominated by 
recycled gas below $z\!<\!1$ for M$_{\rm halo} > 10^{11}$\,\msun, neglecting AGN effects. This is qualitatively 
consistent with our findings and Figure \ref{fig_ratefracs_vs_redshift} for primordial accretion. They further claimed 
that outflows efficiently suppress both cold and hot modes, where these are defined with the standard \tmax criterion, 
excluding all gas ever belonging to a wind. Our definition of primordial is different, in that it allows for recycling 
in satellites prior to incorporation into the MBP. Given this caveat in the comparison, our findings in Fig 
\ref{fig_netrate_vs_redshift} are also qualitatively consistent with this conclusion. 

\cite{fgk11} also considered kinetic wind models, with various constant velocities and mass loading factors, and measured 
accretion rates separated into cold and hot modes based on instantaneous mass fluxes through radial shells together with 
instantaneous gas temperatures. In this respect the analysis differs significantly from Lagrangian definitions of 
accretion and thermal history, and so is not quantitatively comparable. It furthermore prohibits the identification of 
recycled material as in \cite{opp10} and this work. They focused only on flux through the virial shell (accretion onto 
the halo as opposed to the galaxy), and showed galaxy fluxes only for simulations with no winds. Nonetheless, neglecting  
AGN effects, they found that the net cold gas accretion rate onto low-mass haloes was suppressed, while the effect in 
high-mass haloes was negligible. Specifically, for the halo mass range considered here, the decrease is between zero and a 
factor of a few, relatively independent of redshift, which is tension with the changing balance of hot versus cold 
accretion onto halos found in our Figure \ref{fig_coldfrac_4redshifts}, and likely a result of the different wind 
velocity parametrisation.

\cite{vdv11a} considered many simulations with different feedback physics. They found that while the gas accretion rates 
onto haloes was relatively robust against the presence of feedback, the rates onto galaxies themselves depended 
sensitively on stellar winds as well as metal-line cooling. In particular, at our halo mass scale at $z\!=\!2$, there 
was negligible change in the total smooth accretion rate onto galaxies between their reference and no feedback models, 
although this is a sensitive function of mass. However, they also found an order of magnitude decrease in total smooth 
rates introducing either a density dependent wind scaling or thermal AGN feedback. The balance of hot and cold 
smooth accretion onto galaxies at both $z\!=\!2$ and $z\!=\!0$ was found to be fairly insensitive to feedback, in 
qualitative agreement with our findings. Their AGN model did not alter the cold fraction of smooth accretion onto 
haloes at $z\!=\!2$, in disagreement with our Figure \ref{fig_coldfrac_4redshifts}. 
As an extension, \cite{vdv11b} focused on the impact of AGN feedback on inflow and found that it preferentially prevented 
hot mode gas, with high maximum past temperature, from cooling from the halo onto the galaxy. That is, AGN feedback 
reduces hot halo accretion more than cold accretion, based on analysis of all resolved haloes, although this is presumably 
an important effect only in massive haloes above $\simeq 10^{12}$\,\msun. We do not find this differential effect at 
fixed halo mass, although in this work we have focused on lower mass systems not dominated by AGN effects.

\cite{woods14} includes a combined delayed cooling supernova and early stellar feedback model, a numerically distinct 
mechanism for generation of galactic-scale winds from those just discussed. They found strong recycling, as we do. 
Contrasting weaker to stronger stellar feedback, they concluded that the total gas accretion rates did not change with 
strong feedback, while the balance between cold and hot components did. In particular, that the cold component became 
more important. Comparing to our results for smooth accretion, this result is in qualitative disagreement. However, 
this work did not differentiate the merger contribution, and concludes that more cold gas is available for accretion in 
strong feedback runs because more cold gas is present in satellite haloes, not due to any change in accretion processes. 
This could be interpreted as tentative agreement with our conclusion that the fractional contribution of clumpy accretion to 
the net cold rate is larger in the run with feedback (upper right panel of our Figure \ref{fig_ratefracs_vs_redshift}).

Finally, \cite{ubler14} implements a hybrid thermal/kinetic stellar feedback scheme, considering five haloes slightly 
more massive than those discussed herein. They found that strong outflows generate substantially higher raw accretion 
rates, and that recycled material dominates galactic gas accretion at $z\!<\!1$, in good qualitative agreement with our 
results. They also found that the strong feedback case actually increased the rate of ``first'' (our ``primordial'') 
gas accretion, without differentiating between thermal history or different modes. The cause is likely then related to 
the merger contribution, as above, with similar conclusions.

However, all six studies discussed this far have been conducted with the ``classical'' density formulation of smoothed 
particle hydrodynamics \citep[SPH, see][]{spr05}. They therefore agree in their ``no feedback'' cases with \cite{keres09}, 
finding that the cold mode dominates the accretion rates, and/or the star formation rates, of galaxies of all masses, 
particularly so at the higher redshifts of $z\!=\!3$ or $z\!=\!2$. However, we have previously shown that this result 
is incorrect and a consequence of numerical issues with the hydrodynamical method of SPH \citep{nelson13}. Moreover, 
as typically employed in galaxy formation simulations, SPH is not strictly numerically convergent \citep{zhu14}. As a 
result, it is somewhat unclear how to interpret any subsequent conclusions based on the inclusion of further baryonic 
feedback processes.

\cite{dubois12b} investigated AGN feedback at high redshift ($z\!=\!6$) in a single massive halo ($10^{15}$\,\msun\,at 
$z\!=\!0$), and found that large-scale hot superwinds could morphologically disturb cold filaments and quench cold 
diffuse accretion. This is a substantially different mass scale than the one probed here and we cannot make any direct 
comparison.

It has been pointed out that details of the wind interaction, travel extent, IGM enrichment, as well as its recycling 
properties, the balance between inflow and outflow fluxes, and the timescale of reincorporation may all be sensitive to 
physical model inputs as well as numerical details \citep{opp10}. Furthermore, current models designed to incorporate 
feedback and other physical effects from AGN are diverse and relatively unsophisticated, at least as applied in cosmological 
volumes. Finally, different analysis methodologies can make straightforward comparison between simulations impossible. 
While the studies discussed above seem to agree on several qualitative conclusions, it would seem premature to claim any 
strong agreement between simulations as to the impact of feedback on cosmological gas accretion. To do so would require 
identical analysis techniques applied to either large volumes with statistically robust galaxy populations, or identical 
initial conditions of individual haloes, with equivalent physical models and numerical implementations for non-feedback 
physics, including stellar evolution, metals, cooling, star formation, ISM pressurisation, hydrodynamics, and gravity. 
All of the studies reviewed here differ in one or more of these aspects, motivating a more controlled investigation in 
order to resolve this question.


\section{Conclusions} \label{sConclusions}

In this paper we have compared two simulations, realisations of the same initial conditions evolved with the 
moving mesh code {\small AREPO}, using the same methods for both gravity and hydrodynamics. One, which we label as ``no 
feedback'', includes only the simplest baryonic physics -- radiative cooling of a primordial gas, and star formation. 
The other, which we label as ``feedback'', implements the full physics model of the Illustris simulation project, with 
all models and parameters unchanged. Most importantly, the latter includes metals and metal-line cooling from 
enriched gas, stellar feedback resulting in galactic-scale winds, as well as feedback from supermassive black holes. This 
model has been shown to successfully reproduce a number of key stellar observables, in particular the $z\!=\!0$ stellar 
to halo mass relation, and the star formation rate density as a function of redshift, over the full halo mass range 
we consider here. Our aim was to understand how such a comprehensive set of feedback processes, implemented in a full 
cosmological volume, affects the mechanisms by which galaxies acquire their baryonic material. Focusing on the accretion 
of gas by galaxies over cosmic time, we have arrived at four principal conclusions:

\begin{enumerate}
\item
We first consider the accretion rate of material contributing to the net growth of galaxies. We find that the presence 
of feedback strongly suppresses ``smooth mode'' gas -- originating directly from the intergalactic medium, without 
prior incorporation into a satellite galaxy -- at all redshifts. For accretion onto galaxies in $\sim$10$^{11.5}$\,\msun  
\,haloes, the rate of smooth accretion is reduced by a factor of $\sim$\,10 by $z\!=\!1$, increasingly so towards $z\!=\!0$. 
Furthermore, we find that this suppression is independent of the temperature history of newly acquired gas, implying that 
star formation driven galactic winds have marginal impact on the thermal evolution of smoothly accreting material, and that 
the presence of winds does not preferentially prevent material as a function of its past virial heating, or lack thereof. 
In addition to suppressing the net rates, feedback also reduces the raw inflow rates of smooth accretion by a factor of 
$\sim$2, regardless of redshift.

\item
We examine the spatial distribution, temperature, and dynamics of gas in haloes sufficiently massive to exhibit both 
inflowing streams of gas at the virial radius and strong feedback driven outflows arising from the central galaxy. 
Feedback populates the inner halo ($r/r_{\rm vir} \!<\! 0.5$) with a large mass of cold gas with high covering fraction, 
substantially altering the temperature and velocity structure of halo gas at these small radii. At the virial radius, 
however, gas inflow is largely unaffected by the introduction of our fiducial feedback model -- for example, we find 
no notable difference in the temperature distribution of inflowing material, nor the inwards radial mass flux, 
across $r_{\rm vir}$. The spherical covering fraction of inflowing gas at 0.25\,$r_{\rm vir}$ decreases substantially, 
at $z\!=\!2$ from more than 80\% to less than 50\%, while the rates of both inflow and outflow increase, indicative of 
recycling across this boundary.

\item
Comparing the relative contribution of different accretion modes -- smooth, clumpy (merger), stripped and recycled -- 
we find that the fraction of the total net accretion contributed by smooth accretion is lower in the simulation with 
feedback, by roughly a factor of two across all redshifts, and particularly so for \tmax / \tviracc\,$<$\,1 material 
at $z\!<\!1$, which is suppressed in the feedback run to a negligible level. Gas which has been recycled through a wind 
phase prior to its incorporation into a central galaxy makes up a large fraction of total accretion. For gas entering 
the galaxy for the first time, this fraction is $>$50\% at $z\!<\!1$, and between 10\%\,-\,50\% at $z\!>\!2$.

\item
As a measure of the timescale of accretion, we calculate the time difference between the first (highest redshift) 
virial radius crossing and the first incorporation into the galaxy for accreting gas. For smooth accretion, the 
distribution of these ``halo transit times'' is uni-modal and broad. The mean transit time increases in the presence of 
feedback by a factor of $\simeq$\,2\,-\,3, but independent of halo mass. Furthermore, the distribution of these times is 
\textit{also} independent of halo mass, at least from $10^{10}$\,\msun\,to $10^{12}$\,\msun. This holds true in both 
runs, with and without feedback, indicating that the timescale of accretion through the halo does not exhibit any 
sharp transition point in halo mass, above and below which the process of gas accretion proceeds in a fundamentally 
different manner. The full implications for theory concerning the cooling of hot halo gas, such as those commonly used 
in semi-analytical models of galaxy formation, remain an important and intriguing direction for future work.

\end{enumerate}


\textit{Acknowledgments.}
The computations presented in this paper were performed on the Odyssey cluster at Harvard University. 
VS acknowledges support by the European Research Council under ERC-StG grant EXAGAL-308037. 
LH acknowledges support from NASA grant NNX12AC67G and NSF grant AST-1312095.
DN thanks the anonymous referee for many useful comments and suggestions.

\bibliographystyle{mn2e}
\bibliography{refs}

\begin{thebibliography}{}
\makeatletter
\relax
\def\mn@urlcharsother{\let\do\@makeother \do\$\do\&\do\#\do\^\do\_\do\%\do\~}
\def\mn@doi{\begingroup\mn@urlcharsother \@ifnextchar [ {\mn@doi@}
  {\mn@doi@[]}}
\def\mn@doi@[#1]#2{\def\@tempa{#1}\ifx\@tempa\@empty \href
  {http://dx.doi.org/#2} {doi:#2}\else \href {http://dx.doi.org/#2} {#1}\fi
  \endgroup}
\def\mn@eprint#1#2{\mn@eprint@#1:#2::\@nil}
\def\mn@eprint@arXiv#1{\href {http://arxiv.org/abs/#1} {{\tt arXiv:#1}}}
\def\mn@eprint@dblp#1{\href {http://dblp.uni-trier.de/rec/bibtex/#1.xml}
  {dblp:#1}}
\def\mn@eprint@#1:#2:#3:#4\@nil{\def\@tempa {#1}\def\@tempb {#2}\def\@tempc
  {#3}\ifx \@tempc \@empty \let \@tempc \@tempb \let \@tempb \@tempa \fi \ifx
  \@tempb \@empty \def\@tempb {arXiv}\fi \@ifundefined
  {mn@eprint@\@tempb}{\@tempb:\@tempc}{\expandafter \expandafter \csname
  mn@eprint@\@tempb\endcsname \expandafter{\@tempc}}}

\bibitem[\protect\citeauthoryear{{Abadi}, {Navarro}, {Steinmetz}  \&
  {Eke}}{{Abadi} et~al.}{2003}]{abadi03}
{Abadi} M.~G.,  {Navarro} J.~F.,  {Steinmetz} M.,   {Eke} V.~R.,  2003, \mn@doi
  [\apj] {10.1086/375512}, \href
  {http://adsabs.harvard.edu/abs/2003ApJ...591..499A} {591, 499}

\bibitem[\protect\citeauthoryear{{Agertz} \& {Kravtsov}}{{Agertz} \&
  {Kravtsov}}{2014}]{agertz14}
{Agertz} O.,  {Kravtsov} A.~V.,  2014, preprint, \href
  {http://adsabs.harvard.edu/abs/2014arXiv1404.2613A} {} (\mn@eprint {arXiv}
  {1404.2613})

\bibitem[\protect\citeauthoryear{{Agertz}, {Teyssier}  \& {Moore}}{{Agertz}
  et~al.}{2009}]{agertz09}
{Agertz} O.,  {Teyssier} R.,   {Moore} B.,  2009, \mn@doi [\mnras]
  {10.1111/j.1745-3933.2009.00685.x}, \href
  {http://adsabs.harvard.edu/abs/2009MNRAS.397L..64A} {397, L64}

\bibitem[\protect\citeauthoryear{{Agertz}, {Kravtsov}, {Leitner}  \&
  {Gnedin}}{{Agertz} et~al.}{2013}]{agertz13}
{Agertz} O.,  {Kravtsov} A.~V.,  {Leitner} S.~N.,   {Gnedin} N.~Y.,  2013,
  \mn@doi [\apj] {10.1088/0004-637X/770/1/25}, \href
  {http://adsabs.harvard.edu/abs/2013ApJ...770...25A} {770, 25}

\bibitem[\protect\citeauthoryear{{Aumer}, {White}, {Naab}  \&
  {Scannapieco}}{{Aumer} et~al.}{2013}]{aumer13}
{Aumer} M.,  {White} S.~D.~M.,  {Naab} T.,   {Scannapieco} C.,  2013, \mn@doi
  [\mnras] {10.1093/mnras/stt1230}, \href
  {http://adsabs.harvard.edu/abs/2013MNRAS.434.3142A} {434, 3142}

\bibitem[\protect\citeauthoryear{{Bachmann}, {Dolag}, {Hirschmann}, {Almudena
  Prieto}  \& {Remus}}{{Bachmann} et~al.}{2014}]{bachmann14}
{Bachmann} L.~K.,  {Dolag} K.,  {Hirschmann} M.,  {Almudena Prieto} M.,
  {Remus} R.-S.,  2014, preprint, \href
  {http://adsabs.harvard.edu/abs/2014arXiv1409.3221B} {} (\mn@eprint {arXiv}
  {1409.3221})

\bibitem[\protect\citeauthoryear{{Barnes} \& {Hut}}{{Barnes} \&
  {Hut}}{1986}]{bh86}
{Barnes} J.,  {Hut} P.,  1986, \mn@doi [\nat] {10.1038/324446a0}, \href
  {http://adsabs.harvard.edu/abs/1986Natur.324..446B} {324, 446}

\bibitem[\protect\citeauthoryear{{Bauer} \& {Springel}}{{Bauer} \&
  {Springel}}{2012}]{bauer12}
{Bauer} A.,  {Springel} V.,  2012, \mn@doi [\mnras]
  {10.1111/j.1365-2966.2012.21058.x}, \href
  {http://adsabs.harvard.edu/abs/2012MNRAS.tmp.3102B} {p.~3102}

\bibitem[\protect\citeauthoryear{{Bellovary}, {Brooks}, {Volonteri},
  {Governato}, {Quinn}  \& {Wadsley}}{{Bellovary} et~al.}{2013}]{bellovary13}
{Bellovary} J.,  {Brooks} A.,  {Volonteri} M.,  {Governato} F.,  {Quinn} T.,
  {Wadsley} J.,  2013, \mn@doi [\apj] {10.1088/0004-637X/779/2/136}, \href
  {http://adsabs.harvard.edu/abs/2013ApJ...779..136B} {779, 136}

\bibitem[\protect\citeauthoryear{{Birnboim} \& {Dekel}}{{Birnboim} \&
  {Dekel}}{2003}]{bd03}
{Birnboim} Y.,  {Dekel} A.,  2003, \mn@doi [\mnras]
  {10.1046/j.1365-8711.2003.06955.x}, \href
  {http://adsabs.harvard.edu/abs/2003MNRAS.345..349B} {345, 349}

\bibitem[\protect\citeauthoryear{{Booth}, {Agertz}, {Kravtsov}  \&
  {Gnedin}}{{Booth} et~al.}{2013}]{booth13}
{Booth} C.~M.,  {Agertz} O.,  {Kravtsov} A.~V.,   {Gnedin} N.~Y.,  2013,
  \mn@doi [\apjl] {10.1088/2041-8205/777/1/L16}, \href
  {http://adsabs.harvard.edu/abs/2013ApJ...777L..16B} {777, L16}

\bibitem[\protect\citeauthoryear{{Bower}, {Benson}, {Malbon}, {Helly}, {Frenk},
  {Baugh}, {Cole}  \& {Lacey}}{{Bower} et~al.}{2006}]{bower06}
{Bower} R.~G.,  {Benson} A.~J.,  {Malbon} R.,  {Helly} J.~C.,  {Frenk} C.~S.,
  {Baugh} C.~M.,  {Cole} S.,   {Lacey} C.~G.,  2006, \mn@doi [\mnras]
  {10.1111/j.1365-2966.2006.10519.x}, \href
  {http://adsabs.harvard.edu/abs/2006MNRAS.370..645B} {370, 645}

\bibitem[\protect\citeauthoryear{{Brook}, {Stinson}, {Gibson}, {Shen},
  {Macci{\`o}}, {Obreja}, {Wadsley}  \& {Quinn}}{{Brook}
  et~al.}{2014}]{brook14}
{Brook} C.~B.,  {Stinson} G.,  {Gibson} B.~K.,  {Shen} S.,  {Macci{\`o}} A.~V.,
   {Obreja} A.,  {Wadsley} J.,   {Quinn} T.,  2014, \mn@doi [\mnras]
  {10.1093/mnras/stu1406}, \href
  {http://adsabs.harvard.edu/abs/2014MNRAS.443.3809B} {443, 3809}

\bibitem[\protect\citeauthoryear{{Brooks}, {Governato}, {Quinn}, {Brook}  \&
  {Wadsley}}{{Brooks} et~al.}{2009}]{brooks09}
{Brooks} A.~M.,  {Governato} F.,  {Quinn} T.,  {Brook} C.~B.,   {Wadsley} J.,
  2009, \mn@doi [\apj] {10.1088/0004-637X/694/1/396}, \href
  {http://adsabs.harvard.edu/abs/2009ApJ...694..396B} {694, 396}

\bibitem[\protect\citeauthoryear{{Cen}}{{Cen}}{2014}]{cen14}
{Cen} R.,  2014, \mn@doi [\apjl] {10.1088/2041-8205/789/1/L21}, \href
  {http://adsabs.harvard.edu/abs/2014ApJ...789L..21C} {789, L21}

\bibitem[\protect\citeauthoryear{{Ceverino}, {Klypin}, {Klimek},
  {Trujillo-Gomez}, {Churchill}, {Primack}  \& {Dekel}}{{Ceverino}
  et~al.}{2014}]{cev14}
{Ceverino} D.,  {Klypin} A.,  {Klimek} E.~S.,  {Trujillo-Gomez} S.,
  {Churchill} C.~W.,  {Primack} J.,   {Dekel} A.,  2014, \mn@doi [\mnras]
  {10.1093/mnras/stu956}, \href
  {http://adsabs.harvard.edu/abs/2014MNRAS.442.1545C} {442, 1545}

\bibitem[\protect\citeauthoryear{{Costa}, {Sijacki}  \& {Haehnelt}}{{Costa}
  et~al.}{2014}]{costa14}
{Costa} T.,  {Sijacki} D.,   {Haehnelt} M.~G.,  2014, \mn@doi [\mnras]
  {10.1093/mnras/stu1632}, \href
  {http://adsabs.harvard.edu/abs/2014MNRAS.444.2355C} {444, 2355}

\bibitem[\protect\citeauthoryear{{Croton} et~al.,}{{Croton}
  et~al.}{2006}]{croton06}
{Croton} D.~J.,  et~al., 2006, \mn@doi [\mnras]
  {10.1111/j.1365-2966.2005.09675.x}, \href
  {http://adsabs.harvard.edu/abs/2006MNRAS.365...11C} {365, 11}

\bibitem[\protect\citeauthoryear{{Danovich}, {Dekel}, {Hahn}  \&
  {Teyssier}}{{Danovich} et~al.}{2012}]{danovich12}
{Danovich} M.,  {Dekel} A.,  {Hahn} O.,   {Teyssier} R.,  2012, \mn@doi
  [\mnras] {10.1111/j.1365-2966.2012.20751.x}, \href
  {http://adsabs.harvard.edu/abs/2012MNRAS.422.1732D} {422, 1732}

\bibitem[\protect\citeauthoryear{{Danovich}, {Dekel}, {Hahn}, {Ceverino}  \&
  {Primack}}{{Danovich} et~al.}{2014}]{danovich14}
{Danovich} M.,  {Dekel} A.,  {Hahn} O.,  {Ceverino} D.,   {Primack} J.,  2014,
  preprint, \href {http://adsabs.harvard.edu/abs/2014arXiv1407.7129D} {}
  (\mn@eprint {arXiv} {1407.7129})

\bibitem[\protect\citeauthoryear{{Dav{\'e}}, {Finlator}  \&
  {Oppenheimer}}{{Dav{\'e}} et~al.}{2012}]{dave12}
{Dav{\'e}} R.,  {Finlator} K.,   {Oppenheimer} B.~D.,  2012, \mn@doi [\mnras]
  {10.1111/j.1365-2966.2011.20148.x}, \href
  {http://adsabs.harvard.edu/abs/2012MNRAS.421...98D} {421, 98}

\bibitem[\protect\citeauthoryear{{Dekel} \& {Birnboim}}{{Dekel} \&
  {Birnboim}}{2006}]{db06}
{Dekel} A.,  {Birnboim} Y.,  2006, \mn@doi [\mnras]
  {10.1111/j.1365-2966.2006.10145.x}, \href
  {http://adsabs.harvard.edu/abs/2006MNRAS.368....2D} {368, 2}

\bibitem[\protect\citeauthoryear{{Dekel} et~al.,}{{Dekel}
  et~al.}{2009}]{dekel09}
{Dekel} A.,  et~al., 2009, \mn@doi [\nat] {10.1038/nature07648}, \href
  {http://adsabs.harvard.edu/abs/2009Natur.457..451D} {457, 451}

\bibitem[\protect\citeauthoryear{{Dekel}, {Zolotov}, {Tweed}, {Cacciato},
  {Ceverino}  \& {Primack}}{{Dekel} et~al.}{2013}]{dekel13}
{Dekel} A.,  {Zolotov} A.,  {Tweed} D.,  {Cacciato} M.,  {Ceverino} D.,
  {Primack} J.~R.,  2013, \mn@doi [\mnras] {10.1093/mnras/stt1338}, \href
  {http://adsabs.harvard.edu/abs/2013MNRAS.435..999D} {435, 999}

\bibitem[\protect\citeauthoryear{{Dolag}, {Borgani}, {Murante}  \&
  {Springel}}{{Dolag} et~al.}{2009}]{dolag09}
{Dolag} K.,  {Borgani} S.,  {Murante} G.,   {Springel} V.,  2009, \mn@doi
  [\mnras] {10.1111/j.1365-2966.2009.15034.x}, \href
  {http://adsabs.harvard.edu/abs/2009MNRAS.399..497D} {399, 497}

\bibitem[\protect\citeauthoryear{{Dubois}, {Pichon}, {Devriendt}, {Silk},
  {Haehnelt}, {Kimm}  \& {Slyz}}{{Dubois} et~al.}{2012a}]{dubois12b}
{Dubois} Y.,  {Pichon} C.,  {Devriendt} J.,  {Silk} J.,  {Haehnelt} M.,  {Kimm}
  T.,   {Slyz} A.,  2012a, preprint (astro-ph/1206.5838), \href
  {http://adsabs.harvard.edu/abs/2012arXiv1206.5838D} {}

\bibitem[\protect\citeauthoryear{{Dubois}, {Pichon}, {Haehnelt}, {Kimm},
  {Slyz}, {Devriendt}  \& {Pogosyan}}{{Dubois} et~al.}{2012b}]{dubois12a}
{Dubois} Y.,  {Pichon} C.,  {Haehnelt} M.,  {Kimm} T.,  {Slyz} A.,  {Devriendt}
  J.,   {Pogosyan} D.,  2012b, \mn@doi [\mnras]
  {10.1111/j.1365-2966.2012.21160.x}, \href
  {http://adsabs.harvard.edu/abs/2012MNRAS.423.3616D} {423, 3616}

\bibitem[\protect\citeauthoryear{{Dubois} et~al.,}{{Dubois}
  et~al.}{2014}]{dubois14}
{Dubois} Y.,  et~al., 2014, \mn@doi [\mnras] {10.1093/mnras/stu1227}, \href
  {http://adsabs.harvard.edu/abs/2014MNRAS.444.1453D} {444, 1453}

\bibitem[\protect\citeauthoryear{{Faucher-Gigu{\`e}re}, {Lidz}, {Zaldarriaga}
  \& {Hernquist}}{{Faucher-Gigu{\`e}re} et~al.}{2009}]{fg09}
{Faucher-Gigu{\`e}re} C.-A.,  {Lidz} A.,  {Zaldarriaga} M.,   {Hernquist} L.,
  2009, \mn@doi [\apj] {10.1088/0004-637X/703/2/1416}, \href
  {http://adsabs.harvard.edu/abs/2009ApJ...703.1416F} {703, 1416}

\bibitem[\protect\citeauthoryear{{Faucher-Gigu{\`e}re}, {Kere{\v s}}  \&
  {Ma}}{{Faucher-Gigu{\`e}re} et~al.}{2011}]{fgk11}
{Faucher-Gigu{\`e}re} C.-A.,  {Kere{\v s}} D.,   {Ma} C.-P.,  2011, \mn@doi
  [\mnras] {10.1111/j.1365-2966.2011.19457.x}, \href
  {http://adsabs.harvard.edu/abs/2011MNRAS.417.2982F} {417, 2982}

\bibitem[\protect\citeauthoryear{{Faucher-Giguere}, {Hopkins}, {Keres},
  {Muratov}, {Quataert}  \& {Murray}}{{Faucher-Giguere} et~al.}{2014}]{fg14}
{Faucher-Giguere} C.-A.,  {Hopkins} P.~F.,  {Keres} D.,  {Muratov} A.~L.,
  {Quataert} E.,   {Murray} N.,  2014, preprint, \href
  {http://adsabs.harvard.edu/abs/2014arXiv1409.1919F} {} (\mn@eprint {arXiv}
  {1409.1919})

\bibitem[\protect\citeauthoryear{{Feng}, {Di Matteo}, {Croft}  \&
  {Khandai}}{{Feng} et~al.}{2014}]{feng14}
{Feng} Y.,  {Di Matteo} T.,  {Croft} R.,   {Khandai} N.,  2014, \mn@doi
  [\mnras] {10.1093/mnras/stu432}, \href
  {http://adsabs.harvard.edu/abs/2014MNRAS.440.1865F} {440, 1865}

\bibitem[\protect\citeauthoryear{{Gabor} \& {Bournaud}}{{Gabor} \&
  {Bournaud}}{2014}]{gabor14}
{Gabor} J.~M.,  {Bournaud} F.,  2014, \mn@doi [\mnras] {10.1093/mnrasl/slt139},
  \href {http://adsabs.harvard.edu/abs/2014MNRAS.437L..56G} {437, L56}

\bibitem[\protect\citeauthoryear{{Genel}, {Genzel}, {Bouch{\'e}}, {Naab}  \&
  {Sternberg}}{{Genel} et~al.}{2009}]{genel09}
{Genel} S.,  {Genzel} R.,  {Bouch{\'e}} N.,  {Naab} T.,   {Sternberg} A.,
  2009, \mn@doi [\apj] {10.1088/0004-637X/701/2/2002}, \href
  {http://adsabs.harvard.edu/abs/2009ApJ...701.2002G} {701, 2002}

\bibitem[\protect\citeauthoryear{{Genel}, {Dekel}  \& {Cacciato}}{{Genel}
  et~al.}{2012}]{genel12}
{Genel} S.,  {Dekel} A.,   {Cacciato} M.,  2012, \mn@doi [\mnras]
  {10.1111/j.1365-2966.2012.21652.x}, \href
  {http://adsabs.harvard.edu/abs/2012MNRAS.425..788G} {425, 788}

\bibitem[\protect\citeauthoryear{{Genel}, {Vogelsberger}, {Nelson}, {Sijacki},
  {Springel}  \& {Hernquist}}{{Genel} et~al.}{2013}]{genel13}
{Genel} S.,  {Vogelsberger} M.,  {Nelson} D.,  {Sijacki} D.,  {Springel} V.,
  {Hernquist} L.,  2013, \mn@doi [\mnras] {10.1093/mnras/stt1383}, \href
  {http://adsabs.harvard.edu/abs/2013MNRAS.435.1426G} {435, 1426}

\bibitem[\protect\citeauthoryear{{Genel} et~al.,}{{Genel}
  et~al.}{2014}]{genel14}
{Genel} S.,  et~al., 2014, \mn@doi [\mnras] {10.1093/mnras/stu1654}, \href
  {http://adsabs.harvard.edu/abs/2014MNRAS.445..175G} {445, 175}

\bibitem[\protect\citeauthoryear{{G{\'o}rski}, {Hivon}, {Banday}, {Wandelt},
  {Hansen}, {Reinecke}  \& {Bartelmann}}{{G{\'o}rski} et~al.}{2005}]{gorski05}
{G{\'o}rski} K.~M.,  {Hivon} E.,  {Banday} A.~J.,  {Wandelt} B.~D.,  {Hansen}
  F.~K.,  {Reinecke} M.,   {Bartelmann} M.,  2005, \mn@doi [\apj]
  {10.1086/427976}, \href {http://adsabs.harvard.edu/abs/2005ApJ...622..759G}
  {622, 759}

\bibitem[\protect\citeauthoryear{{Hanasz}, {Lesch}, {Naab}, {Gawryszczak},
  {Kowalik}  \& {W{\'o}lta{\'n}ski}}{{Hanasz} et~al.}{2013}]{hanasz13}
{Hanasz} M.,  {Lesch} H.,  {Naab} T.,  {Gawryszczak} A.,  {Kowalik} K.,
  {W{\'o}lta{\'n}ski} D.,  2013, \mn@doi [\apjl] {10.1088/2041-8205/777/2/L38},
  \href {http://adsabs.harvard.edu/abs/2013ApJ...777L..38H} {777, L38}

\bibitem[\protect\citeauthoryear{{Henriques}, {White}, {Thomas}, {Angulo},
  {Guo}, {Lemson}, {Springel}  \& {Overzier}}{{Henriques}
  et~al.}{2014}]{henriques14}
{Henriques} B.,  {White} S.,  {Thomas} P.,  {Angulo} R.,  {Guo} Q.,  {Lemson}
  G.,  {Springel} V.,   {Overzier} R.,  2014, preprint, \href
  {http://adsabs.harvard.edu/abs/2014arXiv1410.0365H} {} (\mn@eprint {arXiv}
  {1410.0365})

\bibitem[\protect\citeauthoryear{{Hopkins}, {Kere{\v s}}, {O{\~n}orbe},
  {Faucher-Gigu{\`e}re}, {Quataert}, {Murray}  \& {Bullock}}{{Hopkins}
  et~al.}{2014}]{hop14fire}
{Hopkins} P.~F.,  {Kere{\v s}} D.,  {O{\~n}orbe} J.,  {Faucher-Gigu{\`e}re}
  C.-A.,  {Quataert} E.,  {Murray} N.,   {Bullock} J.~S.,  2014, \mn@doi
  [\mnras] {10.1093/mnras/stu1738}, \href
  {http://adsabs.harvard.edu/abs/2014MNRAS.445..581H} {445, 581}

\bibitem[\protect\citeauthoryear{{Karakas}}{{Karakas}}{2010}]{karakas10}
{Karakas} A.~I.,  2010, \mn@doi [\mnras] {10.1111/j.1365-2966.2009.16198.x},
  \href {http://adsabs.harvard.edu/abs/2010MNRAS.403.1413K} {403, 1413}

\bibitem[\protect\citeauthoryear{{Katz}, {Weinberg}  \& {Hernquist}}{{Katz}
  et~al.}{1996}]{katz96}
{Katz} N.,  {Weinberg} D.~H.,   {Hernquist} L.,  1996, \mn@doi [\apjs]
  {10.1086/192305}, \href {http://adsabs.harvard.edu/abs/1996ApJS..105...19K}
  {105, 19}

\bibitem[\protect\citeauthoryear{{Katz}, {Keres}, {Dave}  \& {Weinberg}}{{Katz}
  et~al.}{2003}]{katz03}
{Katz} N.,  {Keres} D.,  {Dave} R.,   {Weinberg} D.~H.,  2003, in {Rosenberg}
  J.~L.,  {Putman} M.~E.,  eds,  Astrophysics and Space Science Library Vol.
  281, The IGM/Galaxy Connection. The Distribution of Baryons at z=0. p.~185
  (\mn@eprint {} {arXiv:astro-ph/0209279})

\bibitem[\protect\citeauthoryear{{Kere{\v s}}, {Katz}, {Weinberg}  \&
  {Dav{\'e}}}{{Kere{\v s}} et~al.}{2005}]{keres05}
{Kere{\v s}} D.,  {Katz} N.,  {Weinberg} D.~H.,   {Dav{\'e}} R.,  2005, \mn@doi
  [\mnras] {10.1111/j.1365-2966.2005.09451.x}, \href
  {http://adsabs.harvard.edu/abs/2005MNRAS.363....2K} {363, 2}

\bibitem[\protect\citeauthoryear{{Kere{\v s}}, {Katz}, {Fardal}, {Dav{\'e}}  \&
  {Weinberg}}{{Kere{\v s}} et~al.}{2009}]{keres09}
{Kere{\v s}} D.,  {Katz} N.,  {Fardal} M.,  {Dav{\'e}} R.,   {Weinberg} D.~H.,
  2009, \mn@doi [\mnras] {10.1111/j.1365-2966.2009.14541.x}, \href
  {http://adsabs.harvard.edu/abs/2009MNRAS.395..160K} {395, 160}

\bibitem[\protect\citeauthoryear{{Kere{\v s}}, {Vogelsberger}, {Sijacki},
  {Springel}  \& {Hernquist}}{{Kere{\v s}} et~al.}{2012}]{keres12}
{Kere{\v s}} D.,  {Vogelsberger} M.,  {Sijacki} D.,  {Springel} V.,
  {Hernquist} L.,  2012, \mn@doi [\mnras] {10.1111/j.1365-2966.2012.21548.x},
  \href {http://adsabs.harvard.edu/abs/2012MNRAS.425.2027K} {425, 2027}

\bibitem[\protect\citeauthoryear{{Khandai}, {Di Matteo}, {Croft}, {Wilkins},
  {Feng}, {Tucker}, {DeGraf}  \& {Liu}}{{Khandai} et~al.}{2014}]{khandai14}
{Khandai} N.,  {Di Matteo} T.,  {Croft} R.,  {Wilkins} S.~M.,  {Feng} Y.,
  {Tucker} E.,  {DeGraf} C.,   {Liu} M.-S.,  2014, preprint, \href
  {http://adsabs.harvard.edu/abs/2014arXiv1402.0888K} {} (\mn@eprint {arXiv}
  {1402.0888})

\bibitem[\protect\citeauthoryear{{Li} \& {Bryan}}{{Li} \& {Bryan}}{2014}]{li14}
{Li} Y.,  {Bryan} G.~L.,  2014, \mn@doi [\apj] {10.1088/0004-637X/789/1/54},
  \href {http://adsabs.harvard.edu/abs/2014ApJ...789...54L} {789, 54}

\bibitem[\protect\citeauthoryear{{Lu}, {Kere{\v s}}, {Katz}, {Mo}, {Fardal}  \&
  {Weinberg}}{{Lu} et~al.}{2011}]{lu11}
{Lu} Y.,  {Kere{\v s}} D.,  {Katz} N.,  {Mo} H.~J.,  {Fardal} M.,   {Weinberg}
  M.~D.,  2011, \mn@doi [\mnras] {10.1111/j.1365-2966.2011.19072.x}, \href
  {http://adsabs.harvard.edu/abs/2011MNRAS.416..660L} {416, 660}

\bibitem[\protect\citeauthoryear{{Monaco}, {Benson}, {De Lucia}, {Fontanot},
  {Borgani}  \& {Boylan-Kolchin}}{{Monaco} et~al.}{2014}]{monaco14}
{Monaco} P.,  {Benson} A.~J.,  {De Lucia} G.,  {Fontanot} F.,  {Borgani} S.,
  {Boylan-Kolchin} M.,  2014, \mn@doi [\mnras] {10.1093/mnras/stu655}, \href
  {http://adsabs.harvard.edu/abs/2014MNRAS.441.2058M} {441, 2058}

\bibitem[\protect\citeauthoryear{{Murante}, {Calabrese}, {De Lucia}, {Monaco},
  {Borgani}  \& {Dolag}}{{Murante} et~al.}{2012}]{murante12}
{Murante} G.,  {Calabrese} M.,  {De Lucia} G.,  {Monaco} P.,  {Borgani} S.,
  {Dolag} K.,  2012, \mn@doi [\apjl] {10.1088/2041-8205/749/2/L34}, \href
  {http://adsabs.harvard.edu/abs/2012ApJ...749L..34M} {749, L34}

\bibitem[\protect\citeauthoryear{{Nelson}, {Vogelsberger}, {Genel}, {Sijacki},
  {Kere{\v s}}, {Springel}  \& {Hernquist}}{{Nelson} et~al.}{2013}]{nelson13}
{Nelson} D.,  {Vogelsberger} M.,  {Genel} S.,  {Sijacki} D.,  {Kere{\v s}} D.,
  {Springel} V.,   {Hernquist} L.,  2013, \mn@doi [\mnras]
  {10.1093/mnras/sts595}, \href
  {http://adsabs.harvard.edu/abs/2013MNRAS.429.3353N} {429, 3353}

\bibitem[\protect\citeauthoryear{{Ocvirk}, {Pichon}  \& {Teyssier}}{{Ocvirk}
  et~al.}{2008}]{ocvirk08}
{Ocvirk} P.,  {Pichon} C.,   {Teyssier} R.,  2008, \mn@doi [\mnras]
  {10.1111/j.1365-2966.2008.13763.x}, \href
  {http://adsabs.harvard.edu/abs/2008MNRAS.390.1326O} {390, 1326}

\bibitem[\protect\citeauthoryear{{Oppenheimer}, {Dav{\'e}}, {Kere{\v s}},
  {Fardal}, {Katz}, {Kollmeier}  \& {Weinberg}}{{Oppenheimer}
  et~al.}{2010}]{opp10}
{Oppenheimer} B.~D.,  {Dav{\'e}} R.,  {Kere{\v s}} D.,  {Fardal} M.,  {Katz}
  N.,  {Kollmeier} J.~A.,   {Weinberg} D.~H.,  2010, \mn@doi [\mnras]
  {10.1111/j.1365-2966.2010.16872.x}, \href
  {http://adsabs.harvard.edu/abs/2010MNRAS.406.2325O} {406, 2325}

\bibitem[\protect\citeauthoryear{{Portinari}, {Chiosi}  \&
  {Bressan}}{{Portinari} et~al.}{1998}]{portinari98}
{Portinari} L.,  {Chiosi} C.,   {Bressan} A.,  1998, \aap, \href
  {http://adsabs.harvard.edu/abs/1998A%26A...334..505P} {334, 505}

\bibitem[\protect\citeauthoryear{{Rahmati}, {Pawlik}, {Rai{\v c}evic}  \&
  {Schaye}}{{Rahmati} et~al.}{2013}]{rahmati13}
{Rahmati} A.,  {Pawlik} A.~H.,  {Rai{\v c}evic} M.,   {Schaye} J.,  2013,
  \mn@doi [\mnras] {10.1093/mnras/stt066}, \href
  {http://adsabs.harvard.edu/abs/2013MNRAS.430.2427R} {430, 2427}

\bibitem[\protect\citeauthoryear{{Rees} \& {Ostriker}}{{Rees} \&
  {Ostriker}}{1977}]{rees77}
{Rees} M.~J.,  {Ostriker} J.~P.,  1977, \mnras, \href
  {http://adsabs.harvard.edu/abs/1977MNRAS.179..541R} {179, 541}

\bibitem[\protect\citeauthoryear{{Salem} \& {Bryan}}{{Salem} \&
  {Bryan}}{2014}]{salem14}
{Salem} M.,  {Bryan} G.~L.,  2014, \mn@doi [\mnras] {10.1093/mnras/stt2121},
  \href {http://adsabs.harvard.edu/abs/2014MNRAS.437.3312S} {437, 3312}

\bibitem[\protect\citeauthoryear{{Sales}, {Navarro}, {Theuns}, {Schaye},
  {White}, {Frenk}, {Crain}  \& {Dalla Vecchia}}{{Sales}
  et~al.}{2012}]{sales12}
{Sales} L.~V.,  {Navarro} J.~F.,  {Theuns} T.,  {Schaye} J.,  {White} S.~D.~M.,
   {Frenk} C.~S.,  {Crain} R.~A.,   {Dalla Vecchia} C.,  2012, \mn@doi [\mnras]
  {10.1111/j.1365-2966.2012.20975.x}, \href
  {http://adsabs.harvard.edu/abs/2012MNRAS.423.1544S} {423, 1544}

\bibitem[\protect\citeauthoryear{{S{\'a}nchez Almeida}, {Elmegreen},
  {Mu{\~n}oz-Tu{\~n}{\'o}n}  \& {Elmegreen}}{{S{\'a}nchez Almeida}
  et~al.}{2014}]{almeida14}
{S{\'a}nchez Almeida} J.,  {Elmegreen} B.~G.,  {Mu{\~n}oz-Tu{\~n}{\'o}n} C.,
  {Elmegreen} D.~M.,  2014, \mn@doi [\aapr] {10.1007/s00159-014-0071-1}, \href
  {http://adsabs.harvard.edu/abs/2014A%26ARv..22...71S} {22, 71}

\bibitem[\protect\citeauthoryear{{Sarkar}, {Nath}, {Sharma}  \&
  {Shchekinov}}{{Sarkar} et~al.}{2014}]{sarkar14}
{Sarkar} K.~C.,  {Nath} B.~B.,  {Sharma} P.,   {Shchekinov} Y.,  2014,
  preprint, \href {http://adsabs.harvard.edu/abs/2014arXiv1409.4874S} {}
  (\mn@eprint {arXiv} {1409.4874})

\bibitem[\protect\citeauthoryear{{Schaye} et~al.,}{{Schaye}
  et~al.}{2014}]{schaye14}
{Schaye} J.,  et~al., 2014, preprint, \href
  {http://adsabs.harvard.edu/abs/2014arXiv1407.7040S} {} (\mn@eprint {arXiv}
  {1407.7040})

\bibitem[\protect\citeauthoryear{{Sijacki}, {Springel}, {Di Matteo}  \&
  {Hernquist}}{{Sijacki} et~al.}{2007}]{sijacki07}
{Sijacki} D.,  {Springel} V.,  {Di Matteo} T.,   {Hernquist} L.,  2007, \mn@doi
  [\mnras] {10.1111/j.1365-2966.2007.12153.x}, \href
  {http://adsabs.harvard.edu/abs/2007MNRAS.380..877S} {380, 877}

\bibitem[\protect\citeauthoryear{{Sijacki}, {Vogelsberger}, {Kere{\v s}},
  {Springel}  \& {Hernquist}}{{Sijacki} et~al.}{2012}]{sijacki12}
{Sijacki} D.,  {Vogelsberger} M.,  {Kere{\v s}} D.,  {Springel} V.,
  {Hernquist} L.,  2012, \mn@doi [\mnras] {10.1111/j.1365-2966.2012.21466.x},
  \href {http://adsabs.harvard.edu/abs/2012MNRAS.424.2999S} {424, 2999}

\bibitem[\protect\citeauthoryear{{Silk}}{{Silk}}{1977}]{silk77}
{Silk} J.,  1977, \mn@doi [\apj] {10.1086/154972}, \href
  {http://adsabs.harvard.edu/abs/1977ApJ...211..638S} {211, 638}

\bibitem[\protect\citeauthoryear{{Springel}}{{Springel}}{2010}]{spr10}
{Springel} V.,  2010, \mn@doi [\mnras] {10.1111/j.1365-2966.2009.15715.x}, 401,
  791

\bibitem[\protect\citeauthoryear{{Springel} \& {Hernquist}}{{Springel} \&
  {Hernquist}}{2003}]{spr03}
{Springel} V.,  {Hernquist} L.,  2003, \mn@doi [\mnras]
  {10.1046/j.1365-8711.2003.06206.x}, 339, 289

\bibitem[\protect\citeauthoryear{{Springel}, {White}, {Tormen}  \&
  {Kauffmann}}{{Springel} et~al.}{2001}]{spr01}
{Springel} V.,  {White} S.~D.~M.,  {Tormen} G.,   {Kauffmann} G.,  2001,
  \mn@doi [\mnras] {10.1046/j.1365-8711.2001.04912.x}, \href
  {http://adsabs.harvard.edu/abs/2001MNRAS.328..726S} {328, 726}

\bibitem[\protect\citeauthoryear{{Springel}, {Di Matteo}  \&
  {Hernquist}}{{Springel} et~al.}{2005}]{spr05}
{Springel} V.,  {Di Matteo} T.,   {Hernquist} L.,  2005, \mn@doi [\mnras]
  {10.1111/j.1365-2966.2005.09238.x}, 361, 776

\bibitem[\protect\citeauthoryear{{Stewart}, {Kaufmann}, {Bullock}, {Barton},
  {Maller}, {Diemand}  \& {Wadsley}}{{Stewart} et~al.}{2011}]{stewart11}
{Stewart} K.~R.,  {Kaufmann} T.,  {Bullock} J.~S.,  {Barton} E.~J.,  {Maller}
  A.~H.,  {Diemand} J.,   {Wadsley} J.,  2011, \mn@doi [\apj]
  {10.1088/0004-637X/738/1/39}, \href
  {http://adsabs.harvard.edu/abs/2011ApJ...738...39S} {738, 39}

\bibitem[\protect\citeauthoryear{{Stewart}, {Brooks}, {Bullock}, {Maller},
  {Diemand}, {Wadsley}  \& {Moustakas}}{{Stewart} et~al.}{2013}]{stewart13}
{Stewart} K.~R.,  {Brooks} A.~M.,  {Bullock} J.~S.,  {Maller} A.~H.,  {Diemand}
  J.,  {Wadsley} J.,   {Moustakas} L.~A.,  2013, \mn@doi [\apj]
  {10.1088/0004-637X/769/1/74}, \href
  {http://adsabs.harvard.edu/abs/2013ApJ...769...74S} {769, 74}

\bibitem[\protect\citeauthoryear{{Stinson}, {Brook}, {Macci{\`o}}, {Wadsley},
  {Quinn}  \& {Couchman}}{{Stinson} et~al.}{2013}]{stinson13}
{Stinson} G.~S.,  {Brook} C.,  {Macci{\`o}} A.~V.,  {Wadsley} J.,  {Quinn}
  T.~R.,   {Couchman} H.~M.~P.,  2013, \mn@doi [\mnras] {10.1093/mnras/sts028},
  \href {http://adsabs.harvard.edu/abs/2013MNRAS.428..129S} {428, 129}

\bibitem[\protect\citeauthoryear{{Sutherland} \& {Dopita}}{{Sutherland} \&
  {Dopita}}{1993}]{sd93}
{Sutherland} R.~S.,  {Dopita} M.~A.,  1993, \mn@doi [\apjs] {10.1086/191823},
  \href {http://adsabs.harvard.edu/abs/1993ApJS...88..253S} {88, 253}

\bibitem[\protect\citeauthoryear{{Thielemann}, {Nomoto}  \&
  {Yokoi}}{{Thielemann} et~al.}{1986}]{thiel86}
{Thielemann} F.-K.,  {Nomoto} K.,   {Yokoi} K.,  1986, \aap, \href
  {http://adsabs.harvard.edu/abs/1986A%26A...158...17T} {158, 17}

\bibitem[\protect\citeauthoryear{{Torrey}, {Vogelsberger}, {Sijacki},
  {Springel}  \& {Hernquist}}{{Torrey} et~al.}{2012}]{torrey12}
{Torrey} P.,  {Vogelsberger} M.,  {Sijacki} D.,  {Springel} V.,   {Hernquist}
  L.,  2012, \mn@doi [\mnras] {10.1111/j.1365-2966.2012.22082.x}, \href
  {http://adsabs.harvard.edu/abs/2012MNRAS.427.2224T} {427, 2224}

\bibitem[\protect\citeauthoryear{{Torrey}, {Vogelsberger}, {Genel}, {Sijacki},
  {Springel}  \& {Hernquist}}{{Torrey} et~al.}{2014}]{torrey14}
{Torrey} P.,  {Vogelsberger} M.,  {Genel} S.,  {Sijacki} D.,  {Springel} V.,
  {Hernquist} L.,  2014, \mn@doi [\mnras] {10.1093/mnras/stt2295}, \href
  {http://adsabs.harvard.edu/abs/2014MNRAS.438.1985T} {438, 1985}

\bibitem[\protect\citeauthoryear{{{\"U}bler}, {Naab}, {Oser}, {Aumer}, {Sales}
  \& {White}}{{{\"U}bler} et~al.}{2014}]{ubler14}
{{\"U}bler} H.,  {Naab} T.,  {Oser} L.,  {Aumer} M.,  {Sales} L.~V.,   {White}
  S.~D.~M.,  2014, \mn@doi [\mnras] {10.1093/mnras/stu1275}, \href
  {http://adsabs.harvard.edu/abs/2014MNRAS.443.2092U} {443, 2092}

\bibitem[\protect\citeauthoryear{{Viola}, {Monaco}, {Borgani}, {Murante}  \&
  {Tornatore}}{{Viola} et~al.}{2008}]{viola08}
{Viola} M.,  {Monaco} P.,  {Borgani} S.,  {Murante} G.,   {Tornatore} L.,
  2008, \mn@doi [\mnras] {10.1111/j.1365-2966.2007.12598.x}, \href
  {http://adsabs.harvard.edu/abs/2008MNRAS.383..777V} {383, 777}

\bibitem[\protect\citeauthoryear{{Vogelsberger}, {Sijacki}, {Kere{\v s}},
  {Springel}  \& {Hernquist}}{{Vogelsberger} et~al.}{2012}]{vog12}
{Vogelsberger} M.,  {Sijacki} D.,  {Kere{\v s}} D.,  {Springel} V.,
  {Hernquist} L.,  2012, \mn@doi [\mnras] {10.1111/j.1365-2966.2012.21590.x},
  \href {http://adsabs.harvard.edu/abs/2012MNRAS.425.3024V} {425, 3024}

\bibitem[\protect\citeauthoryear{{Vogelsberger}, {Genel}, {Sijacki}, {Torrey},
  {Springel}  \& {Hernquist}}{{Vogelsberger} et~al.}{2013}]{vog13}
{Vogelsberger} M.,  {Genel} S.,  {Sijacki} D.,  {Torrey} P.,  {Springel} V.,
  {Hernquist} L.,  2013, \mn@doi [\mnras] {10.1093/mnras/stt1789}, \href
  {http://adsabs.harvard.edu/abs/2013MNRAS.436.3031V} {436, 3031}

\bibitem[\protect\citeauthoryear{{Vogelsberger} et~al.,}{{Vogelsberger}
  et~al.}{2014a}]{vog14b}
{Vogelsberger} M.,  et~al., 2014a, \mn@doi [\mnras] {10.1093/mnras/stu1536},
  \href {http://adsabs.harvard.edu/abs/2014MNRAS.444.1518V} {444, 1518}

\bibitem[\protect\citeauthoryear{{Vogelsberger} et~al.,}{{Vogelsberger}
  et~al.}{2014b}]{vog14a}
{Vogelsberger} M.,  et~al., 2014b, \mn@doi [\nat] {10.1038/nature13316}, \href
  {http://adsabs.harvard.edu/abs/2014Natur.509..177V} {509, 177}

\bibitem[\protect\citeauthoryear{{White} \& {Frenk}}{{White} \&
  {Frenk}}{1991}]{wf91}
{White} S.~D.~M.,  {Frenk} C.~S.,  1991, \mn@doi [\apj] {10.1086/170483}, \href
  {http://adsabs.harvard.edu/abs/1991ApJ...379...52W} {379, 52}

\bibitem[\protect\citeauthoryear{{White} \& {Rees}}{{White} \&
  {Rees}}{1978}]{wr78}
{White} S.~D.~M.,  {Rees} M.~J.,  1978, \mnras, \href
  {http://adsabs.harvard.edu/abs/1978MNRAS.183..341W} {183, 341}

\bibitem[\protect\citeauthoryear{{Wiersma}, {Schaye}  \& {Smith}}{{Wiersma}
  et~al.}{2009}]{wiersma09}
{Wiersma} R.~P.~C.,  {Schaye} J.,   {Smith} B.~D.,  2009, \mn@doi [\mnras]
  {10.1111/j.1365-2966.2008.14191.x}, \href
  {http://adsabs.harvard.edu/abs/2009MNRAS.393...99W} {393, 99}

\bibitem[\protect\citeauthoryear{{Woods}, {Wadsley}, {Couchman}, {Stinson}  \&
  {Shen}}{{Woods} et~al.}{2014}]{woods14}
{Woods} R.~M.,  {Wadsley} J.,  {Couchman} H.~M.~P.,  {Stinson} G.,   {Shen} S.,
   2014, \mn@doi [\mnras] {10.1093/mnras/stu895}, \href
  {http://adsabs.harvard.edu/abs/2014MNRAS.442..732W} {442, 732}

\bibitem[\protect\citeauthoryear{{Wurster} \& {Thacker}}{{Wurster} \&
  {Thacker}}{2013}]{wurster13}
{Wurster} J.,  {Thacker} R.~J.,  2013, \mn@doi [\mnras] {10.1093/mnras/stt346},
  \href {http://adsabs.harvard.edu/abs/2013MNRAS.431.2513W} {431, 2513}

\bibitem[\protect\citeauthoryear{{Yoshida}, {Stoehr}, {Springel}  \&
  {White}}{{Yoshida} et~al.}{2002}]{yoshida02}
{Yoshida} N.,  {Stoehr} F.,  {Springel} V.,   {White} S.~D.~M.,  2002, \mn@doi
  [\mnras] {10.1046/j.1365-8711.2002.05661.x}, \href
  {http://adsabs.harvard.edu/abs/2002MNRAS.335..762Y} {335, 762}

\bibitem[\protect\citeauthoryear{{Zhu}, {Hernquist}  \& {Li}}{{Zhu}
  et~al.}{2014}]{zhu14}
{Zhu} Q.,  {Hernquist} L.,   {Li} Y.,  2014, preprint, \href
  {http://adsabs.harvard.edu/abs/2014arXiv1410.4222Z} {} (\mn@eprint {arXiv}
  {1410.4222})

\bibitem[\protect\citeauthoryear{{van Leer}}{{van Leer}}{1977}]{vl77}
{van Leer} B.,  1977, \mn@doi [Journal of Computational Physics]
  {10.1016/0021-9991(77)90095-X}, \href
  {http://adsabs.harvard.edu/abs/1977JCoPh..23..276V} {23, 276}

\bibitem[\protect\citeauthoryear{{van de Voort}, {Schaye}, {Booth}, {Haas}  \&
  {Dalla Vecchia}}{{van de Voort} et~al.}{2011a}]{vdv11a}
{van de Voort} F.,  {Schaye} J.,  {Booth} C.~M.,  {Haas} M.~R.,   {Dalla
  Vecchia} C.,  2011a, \mn@doi [\mnras] {10.1111/j.1365-2966.2011.18565.x},
  \href {http://adsabs.harvard.edu/abs/2011MNRAS.414.2458V} {414, 2458}

\bibitem[\protect\citeauthoryear{{van de Voort}, {Schaye}, {Booth}  \& {Dalla
  Vecchia}}{{van de Voort} et~al.}{2011b}]{vdv11b}
{van de Voort} F.,  {Schaye} J.,  {Booth} C.~M.,   {Dalla Vecchia} C.,  2011b,
  \mn@doi [\mnras] {10.1111/j.1365-2966.2011.18896.x}, \href
  {http://adsabs.harvard.edu/abs/2011MNRAS.415.2782V} {415, 2782}

\makeatother
\end{thebibliography}

\end{document}